\documentclass[twocolumn,prb,amsmath,amssymb,floatfix,superscriptaddress]{revtex4-2} 
\usepackage{amsmath}
\usepackage{amssymb}
\usepackage{graphicx}
\usepackage{bm}
\usepackage{color}
\usepackage{hyperref}
\usepackage{graphicx}
\usepackage{upgreek}
\usepackage{scalerel}
\usepackage{multirow}
 
 \usepackage[bb=dsserif]{mathalpha}
 
\usepackage{pgffor}
\usepackage{pdfpages}
\usepackage{mathdots}
\usepackage{float}
\usepackage{silence}
\usepackage{physics} 
\usepackage{lscape}   
\usepackage{color, colortbl}
\usepackage[table]{xcolor}
\usepackage{comment}  
\makeatletter
\AtBeginDocument{\let\LS@rot\@undefined}
\makeatother

\begin{document}

\title{Classification of pair symmetries in superconductors with unconventional magnetism}

\author{Kazuki Maeda}
\affiliation{Department of Applied Physics, Nagoya University, 464-8603 Nagoya, Japan}

\author{Yuri Fukaya}
\affiliation{Faculty of Environmental Life, Natural Science and Technology, Okayama University, 700-8530 Okayama, Japan}

\author{Keiji Yada}
\affiliation{Department of Applied Physics, Nagoya University, 464-8603 Nagoya, Japan}

\author{Bo Lu}
\affiliation{Department of Physics, Tianjin University, Tianjin 300354, China}

\author{Yukio Tanaka}
\affiliation{Department of Applied Physics, Nagoya University, 464-8603 Nagoya, Japan}

\author{Jorge Cayao}
\affiliation{Department of Physics and Astronomy, Uppsala University, Box 516, S-751 20 Uppsala, Sweden}
\date{\today}

\begin{abstract}
We consider unconventional magnets  with proximity-induced superconductivity and investigate the emergence of superconducting correlations by carrying out a full classification of allowed Cooper pair symmetries. 
  In particular, we focus on $d$-wave altermagnets and $p$-wave magnets under the influence of spin-singlet and spin-triplet superconductivity induced by proximity effect. Under generic conditions, we find that unconventional magnets not only drive a spin-singlet to spin-triplet conversion but also they transfer their parity symmetry that induces superconducting correlations with higher angular momentum. For instance, a conventional spin-singlet $s$-wave superconductor with $d$-wave altermagnetism is able to host odd-frequency mixed spin-triplet $d$-wave superconducting pair amplitudes, while when combining with $p$-wave magnetism the emerging superconducting pairing acquires an even-frequency mixed spin-triplet $p$-wave symmetry. We further demonstrate that unconventional magnetism produces even more exotic superconducting correlations in spin-singlet $d$-wave superconductors, where odd-frequency mixed spin-triplet $g$-wave and even-frequency mixed spin-triplet $f$-wave pair symmetries are possible in altermagnets and $p$-wave magnets, respectively. We also discuss how these ideas generalize to spin-triplet $p$-wave superconductors and also show how our results can be applied to unconventional magnets with higher angular momentum, such as with $f$-, $g$-, and $i$-wave symmetries. Our results can help understand the emergent superconducting correlations due to the interplay of unconventional magnetism and superconductivity.
\end{abstract}
\maketitle

\section{Introduction}
The interplay between magnetism and superconductivity represents one of the most studied areas in physics due to its  fascinating ground for novel superconducting phenomena that hold potential for quantum applications \cite{
Buzdin,Efetov2,Linder_2015,Eschrig2015,Newhorizons_spintronics}. 
Recently, the field has experienced a new boost due to the prediction of magnets exhibiting zero net magnetization and a momentum dependent spin splitting of energy bands which leads to anisotropic spin-polarized Fermi surfaces \cite{landscape22,MazinPRX22}. 
Interestingly, these unconventional magnets can possess even- and odd-parity magnetic orders originating from nonrelativistic effects and protected by distinct symmetries. On one hand, the even parity magnetic order can have $d$-, $g$-, or $i$-wave symmetries and characterizes the so-called altermagnets (AMs), where time-reversal symmetry is broken but inversion symmetry is preserved. 
 The candidate materials for AMs include   RuO$_{2}$ \cite{LiborPRX22,Ahn19}, Mn$_{5}$Si$_{3}$ \cite{LiborPRX22,landscape22}, and   MnF$_{2}$ \cite{Moreno16}. On the other hand, the odd-parity magnetic order appears with $p$-wave symmetry and defines the unconventional $p$-wave magnets (UPMs), where time-reversal symmetry is preserved but inversion symmetry is broken. 
  In terms of candidate materials, UPMs can be realized in Mn$_3$GaN and CeNiAsO \cite{Libor011028}. It is worth noting that higher odd-parity magnetic order is also possible, with $f$-wave magnets being one type of such order \cite{hellenes2024P,jungwirth2024H,ezawa2024}.

The intriguing properties of AMs and UPMs have motivated several studies in the normal state and recently also in systems with superconductivity \cite{FukayaCayaoReview2025},  unveiling unconventional magnetism as a promising ground for realizing exotic superconducting states. In this regard, the combination of superconductivity and  AMs  has already proven very fruitful, with studies including  anomalous Andreev reflections \cite{maeda2024,Papaj23,Sun23,nagae2024,Bo2025}, exotic Josephson effects \cite{Ouassou23,Beenakker23,Cheng24,Bo2024,fukaya2024,sun2024,Bo2025,zhao2025Lu}, topological superconductivity \cite{PhysRevB.108.184505,PhysRevLett.133.106601,CCLiu1,CCLiu2,Mondal24}, superconducting diodes \cite{PhysRevB.110.014518,Banerjee24,chakraborty2024perfe},  magnetoelectric effects  \cite{zyuzin2024,HU2024non}, quantum transport \cite{kokkeler2024quantumt}, {and finite-momentum superconductivity with zero net magnetization \cite{zhang2024,chakraborty2024,sim2024,hong2024,mukasa2024}}.
In relation to UPMs, their combination with superconductors has received less attention and the few studies address  Andreev reflections \cite{maeda2024}, the Josephson effect  \cite{fukaya2024}, and transport \cite{kokkeler2024quantumt}. While all these efforts indeed reveal the great potential of unconventional magnetism,  it is still not fully understood what types of superconducting states emerge due to the interplay of superconductivity and unconventional magnetism.

In this work we consider unconventional magnets with distinct types of superconductivity,   induced  e. g., by proximity effect, and explore how their interplay impacts the symmetries of the emergent superconducting correlations. For this purpose, we address $d$-wave AMs and $p$-wave magnets with spin-singlet and spin-triplet superconductivity, and then we carry out a full classification of superconducting symmetries. We find that unconventional magnets act as a spin-mixer and induce spin-singlet to spin-triplet conversion in spin-singlet and also in chiral $p$-wave superconductors, while they do not for helical $p$-wave superconductors because the latter involve equal spin Cooper pairs. Interestingly, we also discover that unconventional magnets transfer their parity symmetry, which then gives rise to higher angular momentum emergent superconducting correlations in spin-singlet and spin-triplet superconductors. As such, we find that odd-frequency mixed spin-triplet $d$-wave  superconducting pairing emerges in conventional spin-singlet $s$-wave superconductors with $d$-wave altermagnetism, while even-frequency mixed spin-triplet $p$-wave pairs are possible when combining a conventional superconductor with $p$-wave magnetism. For spin-singlet $d$-wave superconductors combined with   AMs and $p$-wave magnets, we show that odd-frequency mixed spin-triplet $g$-wave and even-frequency mixed spin-triplet $f$-wave superconducting correlations appear, respectively. We obtain a similar spin and parity conversion   in chiral spin-triplet $p$-wave superconductors, with an intriguing odd-frequency spin-singlet $f$-wave pairing in $d$-wave AMs. Furthermore, we demonstrate that our findings can be easily used for unconventional magnets with higher angular momentum, such as with $f$-, $g$-, and $i$-wave symmetries, which then produce superconducting correlations with an even higher degree in momentum and might help conceive previously reported predicted higher angular momentum superconducting states. \cite{Kuroki2001, TanakaKuroki2004, Kuroki2004, Kuroki2005}.

The remainder of this work is organized as follows. In Section~\ref{section2} we discuss the models for unconventional magnets, the distinct types of superconductors,  the method for obtaining the superconducting correlations, and density of states (DOS). In  Section~\ref{section3} we discuss the superconducting correlations in superconductors with $d$-wave AMs, while in Section~\ref{section4} we focus on superconductors with $p$-wave magnets, and we present our conclusions in Section~\ref{section5}.  In  Appendix \ref{AppendixA}, we show how our results can be used for obtaining the superconducting correlations in unconventional magnets with higher angular momentum.

\begin{figure}
  \centering
  \includegraphics[width=0.4\textwidth]{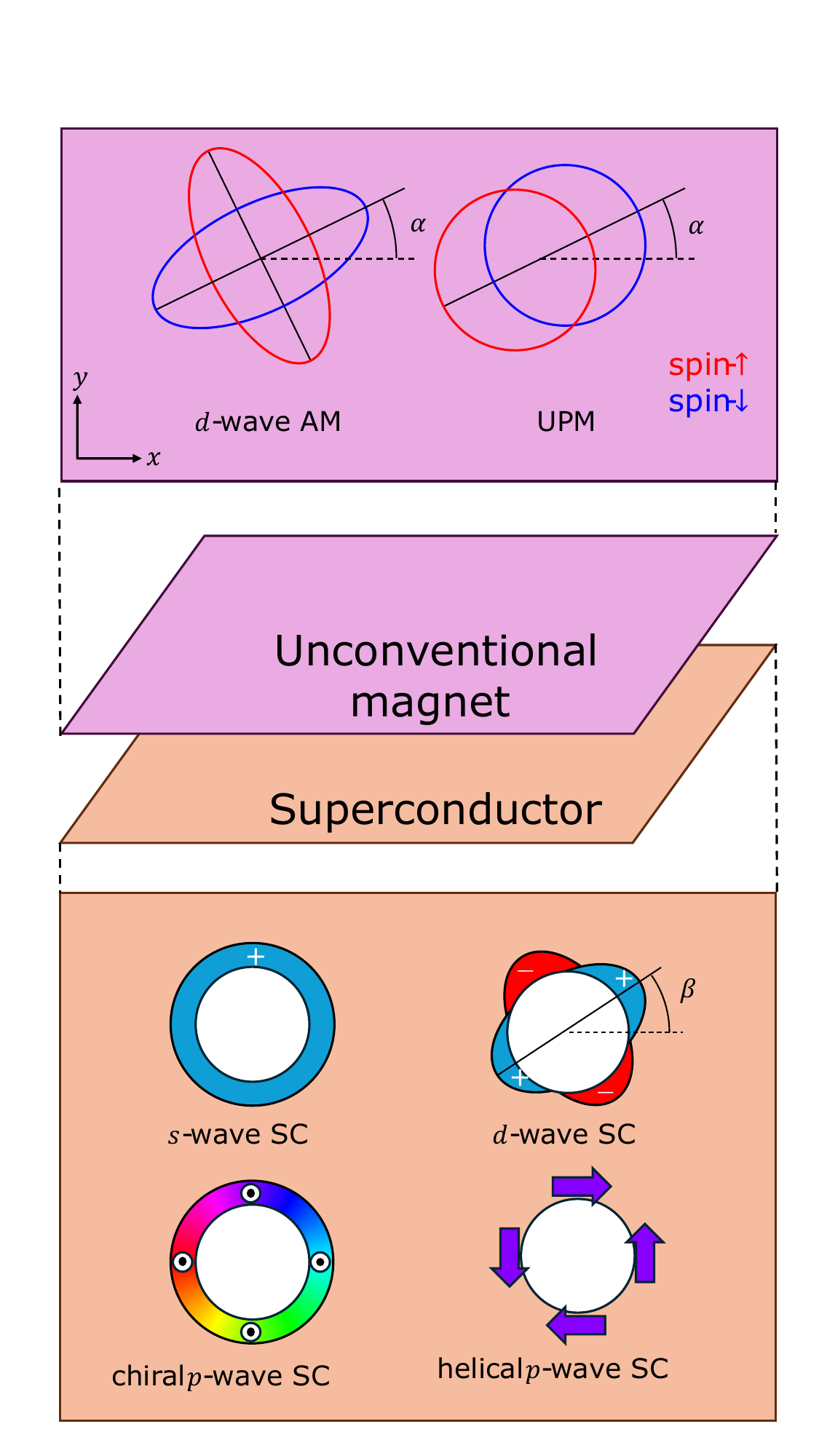}
  \caption{Schematic figure of the interface between a two-dimensional unconventional magnet and a two-dimensional superconductor (middle panel). The top panel shows the shape of the Fermi surfaces of the unconventional magnets, the $d$-wave AM and UPM. The red and blue colors indicate the up and down spins, respectively, while $\alpha$ is the angle between the $x$-axis and the lobe of the unconventional magnet. The bottom panel shows the pair potentials $d_0(\bm{k})$ or $\bm{d}(\bm{k})$ of the distinct types of superconductors. The colors represent the complex argument of the momentum dependence of the pair potentials:    $+$ (cyan) and $-$ (red) signs indicate the positive and negative values, while purple and green colors represent the complex arguments $\pi/2$ and  $-\pi/2$; $\beta$ is the angle between the $x$-axis and the lobe of the $d$-wave superconductor.   The arrows indicate the directions of the $d$-vectors of the spin-triplet superconductors.  The little circles with a black dot inside them represent arrows directed out of the plane. }
  \label{Fig1}
\end{figure}
\section{Model and Method}
\label{section2}
We are interested in investigating the types of superconducting correlations that emerge when combining superconductors and unconventional magnets, as depicted in Fig.\,\ref{Fig1}. We consider a two-dimensional (2D) unconventional magnet modelled by 
\begin{equation}
  H_{N} = \sum_{\bm{k}} \hat{c}_{\bm{k}}^\dagger \hat{h}(\bm{k}) \hat{c}_{\bm{k}},
\end{equation}
with 
\begin{equation}
\label{hN}
  \hat{h}(\bm{k}) = \xi_{\bm{k}} \hat{\sigma}_0 + M_{\bm{k}} \hat{\bm{n}}\cdot\hat{\bm{\sigma}}
  \end{equation}
where $\hat{c}_{\bm{k}}=(c_{\bm{k}\uparrow},c_{\bm{k}\downarrow})^{\top}$, $c_{\bm{k}\sigma}$ is a fermionic operator that annihilates an electronic state with momentum  $\bm{k}=(k_{x},k_{y})$ and spin $\sigma$. Moreover, $\xi_{\bm{k}} = \hbar^2|\bm{k}|^2/(2m) - \mu$ is the kinetic energy, $\mu$ is the chemical potential measuring the filling of the band, $M_{\bm{k}}$ is the strength of the unconventional magnet that depends on the wavenumber $\bm{k}$, $\hat{\bm{n}}$ is the unit vector in the direction of the magnetization, and $\hat{\sigma}_i$ with $i=0,x,y,z$ are the Pauli matrices in the spin space. We focus on recently predicted unconventional magnets, including $d$-wave AMs and unconventional $p$-wave magnets. The $d$-wave AMs are described by  an effective momentum dependent exchange field given by 
  \begin{equation}
  \label{eq:exchange_dAM}
  M^{d}_{\bm{k}} = \frac{J}{k_\mathrm{F}^2}\left[(k_x^2-k_y^2)\cos2\alpha+2k_xk_y\sin2\alpha\right],
\end{equation}
while the unconventional $p$-wave magnets by
\begin{equation}
\label{eq:exchange_UPM}
  M^{p}_{\bm{k}} = \frac{J}{k_\mathrm{F}}\left[k_x\cos\alpha+k_y\sin\alpha\right]\,.
\end{equation}
Here, $J$ is the exchange energy, $k_\mathrm{F}=\sqrt{2m\mu}/\hbar$ is the Fermi wavenumber, and $\alpha$ represents the angle between the $x$-axis and the lobe of AM or unconventional $p$-wave magnet, see Fig.\,\ref{Fig1}(a). We note that the  exchange field strength for AMs is an even function of momentum (even parity), namely, $M^{d}_{\bm{k}}=M^{d}_{-\bm{k}}$. For UPMs, the exchange field obeys $M^{p}_{\bm{k}}=-M^{p}_{-\bm{k}}$, which is an odd function of momentum (odd parity).  The even parity symmetry of AMs also holds for unconventional magnets with a higher even power in momentum such as $g$- and $i$-wave magnets since $M^{g}_{\bm{k}}=J k_{x}k_{y}(k_{x}^{2}-k_{y}^{2})/k_{\rm F}^{4}$ and  $M^{i}_{\bm{k}}=Jk_{x}k_{y}(3k_{x}^{2}-k_{y}^{2})(3k_{y}^{2}-k_{x}^{2})/k_{\rm F}^{6}$. Similarly, the odd parity symmetry is also inherent to unconventional magnets with a higher odd power in momentum such as in $f$-wave magnets, where $M^{f}_{\bm{k}}=J k_{x}(k_{x}^{2}-3k_{y}^{2})/k_{\rm F}^{3}$. While we will mainly focus on AMs and UPMs, whenever necessary we will also discuss unconventional magnets with higher angular momentum dependence.

To account for superconductivity, we assume that the unconventional magnet possesses homogeneously induced superconductivity e.g., by proximity effect, as depicted in Fig.\,\ref{Fig1}.  The unconventional magnet with superconductivity can therefore be described by a mean-field Hamiltonian written in Nambu space as
\begin{equation}\label{eq:Hamiltonian_BdG}
  H_{\rm S} = \frac{1}{2}\sum_{\bm{k}} \Psi_{\bm{k}}^\dagger \check{H}_{\mathrm{BdG}}(\bm{k}) \Psi_{\bm{k}},
\end{equation}
with $\Psi_{\bm{k}}=(c_{\bm{k}\uparrow},c_{\bm{k}\downarrow},c^\dagger_{-\bm{k}\uparrow},c^\dagger_{-\bm{k}\downarrow})^{\top}$ being the Nambu spinor and $ \check{H}_{\mathrm{BdG}}$ the   Bogoliubov-de Gennes (BdG) Hamiltonian
given by
\begin{equation}
\label{HBdG}
  \check{H}_{\mathrm{BdG}}(\bm{k}) = \begin{pmatrix}
    \hat{h}(\bm{k}) & \hat{\Delta}(\bm{k}) \\
    \hat{\Delta}^\dagger(\bm{k}) & -\hat{h}^*(-\bm{k})
  \end{pmatrix},
\end{equation}
where $\hat{\Delta}(\bm{k})$ represents the pair potential and $\hat{h}$ is the normal state Hamiltonian given by Eq.\,(\ref{hN}). Taking into account the spin structure, the pair potential can be written as \cite{SigristUeda}
\begin{equation}
\label{DeltaEq}
  \hat{\Delta}(\bm{k}) = \left[d_0(\bm{k}) \hat{\sigma}_0 + \bm{d}(\bm{k}) \cdot \hat{\bm{\sigma}}\right]i\hat{\sigma}_y,
\end{equation}
with $d_0(\bm{k})$ the spin-singlet pair potential and $\bm{d}(\bm{k})$ the $d$-vector of the spin-triplet pair potential. We note that the pair potential satisfies the relations $d_0(\bm{k}) = d_0(-\bm{k})$ and $\bm{d}(\bm{k}) = -\bm{d}(-\bm{k})$, which can be shown by using the Fermionic commutation relations. We here will analyze distinct types of superconductors, including spin-singlet $s$- and $d$-wave as well as spin-triplet chiral \cite{Leggett,Yamashiro97,Honerkamp1998,Furusaki2001,Maeno2012} and helical $p$-wave superconductors \cite{Frigeri,Iniotakis2007,QiPRL,Tanaka2009}. The spin-singlet   and the  spin-triplet chiral $p$-wave pair potential correspond to the spin projection with $S_{z}=0$, while the spin-triplet helical $p$-wave pair potential to the spin projection with $S_{z}=\pm1$. The specific form of their pair potentials are detailed in Table \ref{tab:pair_potentials}. We expect that the Hamiltonian on the interface between the unconventional magnet and the superconductor in Fig.~\ref{Fig1} can be approximated by Eq.~\eqref{eq:Hamiltonian_BdG}.  It is worth noting that, in terms of materials, conventional spin-singlet $s$-wave superconductors can be fabricated using Al, Nb, or NbTiN \cite{Lutchyn2018}, while spin-singlet $d$-wave superconductors are likely to appear e. g., in cuprates \cite{Kashiwaya_2000}. In the case of spin-triplet $p$-wave superconductors, there already exist evidence of their emergence in several materials, such as in UPt$_3$ \cite{schemm2014observation},    UTe$_2$ \cite{ran2019nearly}, and K$_2$Cr$_3$As$_3$  \cite{yang2021spin}. Alternatively,   spin-triplet $p$-wave pair potentials can be designed by combining a conventional spin-singlet $s$-wave superconductor with spin-orbit coupling  e. g., in Al-InAs systems \cite{Lutchyn2018}.

\begin{table}\label{tab:pair_potentials}
  \centering
  \caption{Expressions of the $d$ vectors describing the pair potentials of the superconductors considered in this work. While $d_{0}(\bm{k})$  describes   the spin-singlet $s$- and $d$-wave superconductors,  $\bm{d}(\bm{k})$ describes the spin-triplet chiral and helical $p$-wave superconductors. The spin-triplet chiral (helical)  $p$-wave pair potential corresponds to the mixed spin-triplet (equal spin-triplet) superconductor \cite{bernevig2013topological}. Here,   $\Delta_0$ is a constant quantity representing the amplitude of the pair potential, while    $\beta$ depicts the angle between the $x$-axis and the lobe of the $d$-wave superconductor \cite{TK95,TK96,tanaka971,Kashiwaya_2000,Lofwander2001}, see  Fig.~\ref{Fig1}.  
  Moreover, $k_{\rm F}$ is the Fermi wavevector and $(\hat{\bm{x}},\hat{\bm{y}},\hat{\bm{z}})$ is a unit vector in 3D.}
  \begin{tabular}{|c|c|}
    \hline
{\bf   Type of superconductor}   & {\bf Pair potential} \\
    \hline\hline
    spin-singlet $s$-wave & $d_0(\bm{k})=\Delta_0$ \\
    spin-singlet $d$-wave & $d_0(\bm{k})=\Delta_0[(k_x^2-k_y^2)\cos2\beta$ \\
    &\quad $+2{k_xk_y}\sin2\beta]/k_\mathrm{F}^2$\\
    spin-triplet chiral $p$-wave & $\bm{d}(\bm{k})=\Delta_0(k_x+ik_y)\hat{\bm{z}}/k_\mathrm{F}$ \\
    spin-triplet helical $p$-wave & $\bm{d}(\bm{k})=i\Delta_0(k_x\hat{\bm{y}}+k_y\hat{\bm{x}})/k_\mathrm{F}$ \\
    \hline
  \end{tabular}
\end{table}

\subsection{Superconducting correlations and density of states}
\label{subsection2a}
To characterize the types of   induced Cooper pairs, we investigate the anomalous electron-hole component of the Gorkov Green's function $\check{G}(\bm{k},i\omega_n)$ in Nambu space.  $\check{G}(\bm{k},i\omega_n)$ is obtained as
\begin{equation}
\label{GorkovG}
  \check{G}(\bm{k},i\omega_n) = \left[i\omega_n\check{1} - \check{H}_{\mathrm{BdG}}(\bm{k})\right]^{-1},
\end{equation}
where $\omega_n$  represents Matsubara frequencies, $\Check{1}$ is the $4\times4$ identity matrix, and $\check{H}_{\mathrm{BdG}}(\bm{k})$ is the BdG Hamiltonian given by Eq.\,(\ref{HBdG}). We then   write $\check{G}(\bm{k},i\omega_n)$ as a  $4\times4$ matrix in Nambu and spin spaces\cite{SigristUeda,zagoskin,TextTanaka2021},
\begin{equation}
  \check{G}(\bm{k},i\omega_n) = 
  \begin{pmatrix}
    \hat{G}_{0}(\bm{k},i\omega_n) & \hat{F}(\bm{k},i\omega_n) \\
    \hat{{\bar{F}}}(\bm{k},i\omega_n) & \hat{{\bar{G}}}_{0}(\bm{k},i\omega_n)
  \end{pmatrix},
\end{equation}
where  $\hat{G}_{0}(\bm{k},i\omega_n)$ and $\hat{{\bar{G}}}_{0}(\bm{k},i\omega_n)$ represent  the normal Green's functions, while $\hat{F}(\bm{k},i\omega_n)$ and $\hat{{\bar{F}}}(\bm{k},i\omega_n)$ are the   anomalous Green's functions. Both the normal and the anomalous Green's functions are matrices in spin space  given by,
\begin{equation}
\begin{split}
  \hat{G}_{0}(\bm{k},i\omega_n)&    =\begin{pmatrix}
        G_{\uparrow\uparrow}(\bm{k},i\omega_n)   & G_{\uparrow\downarrow}(\bm{k},i\omega_n) \\
        G_{\downarrow\uparrow}(\bm{k},i\omega_n) & G_{\downarrow\downarrow}(\bm{k},i\omega_n)
    \end{pmatrix}\,,\\
  \hat{F}(\bm{k},i\omega_n)
&    =\begin{pmatrix}
        F_{\uparrow\uparrow}(\bm{k},i\omega_n)   & F_{\uparrow\downarrow}(\bm{k},i\omega_n) \\
        F_{\downarrow\uparrow}(\bm{k},i\omega_n) & F_{\downarrow\downarrow}(\bm{k},i\omega_n)
    \end{pmatrix}\,.
    \end{split}
\end{equation}
The normal Green's function allows the calculation of the density of states,  while the anomalous Green's function determines the superconducting correlations. In particular, the DOS is obtained as 
\begin{equation}\label{eq:DOS}
    D(E)=\int A(\bm{k},E)\,d^2\bm{k} \,,
\end{equation}
where $A(\bm{k},E)$ is the spectral function obtained as
\begin{equation}\label{eq:Spectrum_Func}
 A(\bm{k},E)=-\frac{1}{\pi} \mathrm{Im}\,\mathrm{tr}\,\hat{G}_0 (\bm{k},E+i\eta)
\end{equation}
and  $\eta$ is an infinitesimal positive number that enables the analytic continuation to real energies.

To identify the type of superconducting correlation, it is necessary to unveil the symmetry of the pair amplitude  $F_{\sigma\sigma'}(\bm{k},i\omega_n)$, see Refs.\cite{zagoskin,TextTanaka2021}. This is possible by exploiting the antisymmetry condition imposed by the Fermi-Dirac statistics  on the pair amplitude under the total exchange of the quantum numbers \cite{Tanaka2012,Cayao2020,triola2020role,TanakaCayaotheory},
\begin{equation}
\label{Fantisymmetry}
F_{\sigma\sigma'}(\bm{k},i\omega_n)=-F_{\sigma'\sigma}(-\bm{k},-i\omega_n)\,.
\end{equation} 
Under the   exchange of individual quantum numbers, the pair amplitude $F_{\sigma\sigma'}(\bm{k},i\omega_n)$ can be either even or odd but the total exchange must pick up a minus sign as dictated by Eq.\,(\ref{Fantisymmetry}).
Without loss of generality, we can decompose the spin symmetry in the pair amplitudes by writing 
\begin{equation}
\label{eq:pair_amplitude}
  \hat{F}(\bm{k},i\omega_n) = \left[F_{\mathrm{s}}(\bm{k},i\omega_n) \hat{\sigma}_0 + \bm{F}_\mathrm{t}(\bm{k},i\omega_n) \cdot \hat{\bm{\sigma}}\right]i\hat{\sigma}_y,
\end{equation}
where $F_{\mathrm{s}}(\bm{k},i\omega_n)$ corresponds to the pair amplitude of spin-singlet Cooper pairs and $\bm{F}_\mathrm{t}(\bm{k},i\omega_n)$ corresponds to a vector containing the   pair amplitudes of spin-triplet Cooper pairs,
\begin{equation}
\label{Fst}
\begin{split}
F_{\mathrm{s}}(\bm{k},i\omega_n) &=\,\frac12(F_{\uparrow\downarrow}-F_{\downarrow\uparrow}),\\
\bm{F}_{\mathrm{t}}(\bm{k},i\omega_n) &=\,
\left(F_{t}^{x},F_{t}^{y},F_{t}^{z}\right)\,,
\end{split}
\end{equation}
with 
\begin{equation}
\begin{split}
F_{t}^{x}&=\frac12(-F_{\uparrow\uparrow}+F_{\downarrow\downarrow})\,,\\
F_{t}^{y}&=\frac{1}{2i}(F_{\uparrow\uparrow}+F_{\downarrow\downarrow})\,,\\
F_{t}^{z}&=\frac12(F_{\uparrow\downarrow}+F_{\downarrow\uparrow})\,.
\end{split}
\end{equation}
For visualization and further understanding purposes, we also define
\begin{equation}
\begin{split}
  \Phi_{\alpha}(\omega_n) &= \int d^2{\bm{k}}\,\bm{F}_{\alpha}(\bm{k},i\omega_n)\, \bm{F}^{\dagger}_{\alpha}(\bm{k},i\omega_n)\,,
\end{split}
\end{equation}
where $\alpha=s,t$ and hence denotes that $\Phi_{\alpha}(\omega_n)$ is due to a singlet or triplet pair amplitude. Note that the structure of the pair amplitude $\hat{F}(\bm{k},i\omega_n)$ in Eq.\,(\ref{eq:pair_amplitude}) is the same as that of the pair potential $\hat{\Delta}(\bm{k})$ in Eq.\,(\ref{DeltaEq}); $\hat{\Delta}(\bm{k})$ represents the order parameter of the parent superconductor, while  $\hat{F}(\bm{k},i\omega_n)$ describes the emergent superconducting pairing. It is worth noting that the normal and anomalous Green's functions are connected by the following relation,  
\begin{equation}
\label{G0F}
    \hat{G}_0(\bm{k},i\omega_n)=
    \hat{F}(\bm{k},i\omega_n)[i\omega_n\hat{\sigma}_0+\hat{h}^*(-\bm{k})]\hat{\Delta}(\bm{k})^{-1}\,,
\end{equation} 
which provides a useful way to explore the impact of emergent pair correlations on the DOS. Eq.\,(\ref{G0F}) is generic and applies to any type of superconductor.  

Before going further, we anticipate the allowed symmetries of the superconducting correlations. By using the antisymmetry condition given by Eq.\,(\ref{Fantisymmetry}),  the pair amplitudes develop four pair symmetry classes: i) even-frequency spin-singlet even-parity (ESE), ii)  even-frequency spin-triplet odd-parity (ETO), iii) odd-frequency spin-singlet odd-parity (OSO) \cite{Balatsky1992,Fuseya,odd3,odd3b,Eschrig2007,PhysRevB.104.094503,PhysRevB.109.205406}, ii)  odd-frequency spin-triplet even-parity (OTE) 
\cite{Berezinskii74,Kirkpatrick91,Coleman1994,Bergeret2001,PhysRevLett.90.117006,Efetov2,odd1,Tanaka2012,Fominov,PhysRevB.96.155426,
LinderRev19,thanos2019,Cayao2020,PhysRevB.101.094506,PhysRevB.101.214507,PhysRevB.106.L100502,Tanaka2021,TanakaCayaotheory,EslamABSMBSOdd2024}. While all these pair symmetry class are in principle allowed, their existence and size are strongly dependent on the system under study  \footnote{
We note that Ref.\,\cite{PhysRevB.111.054520} studied superconducting order parameters within the framework of spin space groups applicable to UMs and hence complements the study we present here to understand superconductivity in UMs.}.
For instance, within this classification, the spin-singlet (spin-triplet) pair potentials shown in Tab.\,\ref{tab:pair_potentials} correspond to ESE (ETO) pair symmetry classes. In what follows, we will use Eq.\,(\ref{GorkovG}), together with Eq.\,(\ref{HBdG}),  in order to identify all the possible allowed superconducting correlations emergent due to the combination of unconventional magnets and superconductors.

\section{Superconductors with $d$-wave magnetism}
\label{section3}
We start by discussing the emergent superconducting correlations in spin-singlet and spin-triplet superconductors with   $d$-wave altermagnetism characterized by the Hamiltonian given by  Eq.~\eqref{eq:exchange_dAM}. We set the direction of the magnetization as $\hat{\bm{n}}=\hat{\bm{z}}$  with $\hat{\bm{z}}$ the unit vector along the $z$-direction. 

\subsection{Spin-singlet superconductors}
For the spin-singlet pair potentials, with ${d}_{0}(\bm{k})$ due to a spin-singlet $s$-wave or spin-singlet $d$-wave superconductor  given in Tab.\,\ref{tab:pair_potentials} and  $\bm{d}(\bm{k})=0$, we analytically obtain the  anomalous Green's function and its components read,
\begin{equation}
\label{singletF}
\begin{split}
  F_{\uparrow\uparrow}(\bm{k},i\omega_n) &= F_{\downarrow\downarrow}(\bm{k},i\omega_n) = 0,\\
  F_{\uparrow\downarrow}(\bm{k},i\omega_n) &= 
  -\frac{d_0(\bm{k})}{
Q_{\bm{k}}(\omega_{n}) + 2i\omega_nM^{d}_{\bm{k}}},\\
  F_{\downarrow\uparrow}(\bm{k},i\omega_n) &= 
  \frac{d_0(\bm{k})}{Q_{\bm{k}}(\omega_{n}) - 2i\omega_n M^{d}_{\bm{k}}}\,,
    \end{split}
\end{equation}
where $Q_{\bm{k}}(\omega_{n})=\omega_n^2+\xi_{\bm{k}}^2 - [M^{d}_{\bm{k}}]^2+|d_0(\bm{k})|^2$ is an even function in both frequency and momentum.  Using Eqs.\,(\ref{singletF}) and Eqs.\,(\ref{Fst}), we obtain spin-singlet and spin-triplet pair amplitudes given by
\begin{equation}
\label{Fstsingletdwave}
\begin{split}
  F_\mathrm{s}(\bm{k},i\omega_n)&=
  -\frac{d_0(\bm{k})Q_{\bm{k}}(\omega_{n})}
  {[Q_{\bm{k}}(\omega_{n})]^2+4[M_{\bm{k}}^{d}]^2\omega_n^2}\,,\\
  \bm{F}_\mathrm{t}(\bm{k},i\omega_n)&=
  \frac{2i\omega_nd_0(\bm{k}) M^{d}_{\bm{k}}\hat{\bm{z}}}
  {[Q_{\bm{k}}(\omega_{n})]^2+4[M_{\bm{k}}^{d}]^2\omega_n^2}\,.
  \end{split}
\end{equation}
By a direct inspection of previous equations, we identify that the spin-singlet pair amplitude $F_\mathrm{s}(\bm{k},i\omega_n)$ has an even-frequency dependence and is even in parity $\bm{k}$ since $d_0(\bm{k})=d_0(-\bm{k})$ and $Q_{\bm{k}}(\omega_{n})=Q_{-\bm{k}}(\omega_{n})$; hence, $F_\mathrm{s}(\bm{k},i\omega_n)$ has ESE   symmetry which fulfills the antisymmetry condition given by Eq.\,(\ref{Fantisymmetry}). In the case of the spin-triplet $\bm{F}_\mathrm{t}(\bm{k},i\omega_n)$, it develops an odd-frequency dependence  and is even in $\bm{k}$ since both $d_0(\bm{k})$ and $M_{k}^{d}$ are  even functions of momentum. The even momentum dependence due to the AM field $M_{k}^{d}$ induces a parity symmetry in $\bm{F}_\mathrm{t}(\bm{k},i\omega_n)$ of $d$-wave nature. Thus, if $d_0(\bm{k})$ corresponds to a spin-singlet $s$-wave superconductor, the parity of  $\bm{F}_\mathrm{t}(\bm{k},i\omega_n)$ is $d$-wave  due to the AM. If, $d_0(\bm{k})$ corresponds to a spin-singlet $d$-wave superconductor,  the parity of  $\bm{F}_\mathrm{t}(\bm{k},i\omega_n)$ is given by the combined action of both the $d$-wave AM and $d$-wave superconductor,  which can be interpreted as an induced $g$-wave parity. We remark that $\bm{F}_\mathrm{t}(\bm{k},i\omega_n)$ is a spin-triplet component along $z$-axis, which characterizes pairing between distinct spins (mixed spin-triplet). Hence, $\bm{F}_\mathrm{t}(\bm{k},i\omega_n)$ is an OTE pair symmetry class that satisfies the antisymmetry condition given by Eq.\,(\ref{Fantisymmetry}). An interesting point to remark here is that $\bm{F}_\mathrm{t}(\bm{k},i\omega_n)$ originates from the $d$-wave AM field $M^{d}_{\bm{k}}$, which here acts as a spin mixing mechanism that induces mixed spin-triplet superconducting correlations along $\hat{\bm{z}}$.  While ESE pairing stems from the parent superconductor, OTE  pairing is an emergent superconducting correlation due to the interplay of $d$-wave magnetism and spin-singlet superconductivity.

\begin{figure}
    \centering
    \includegraphics[width=0.95\linewidth]{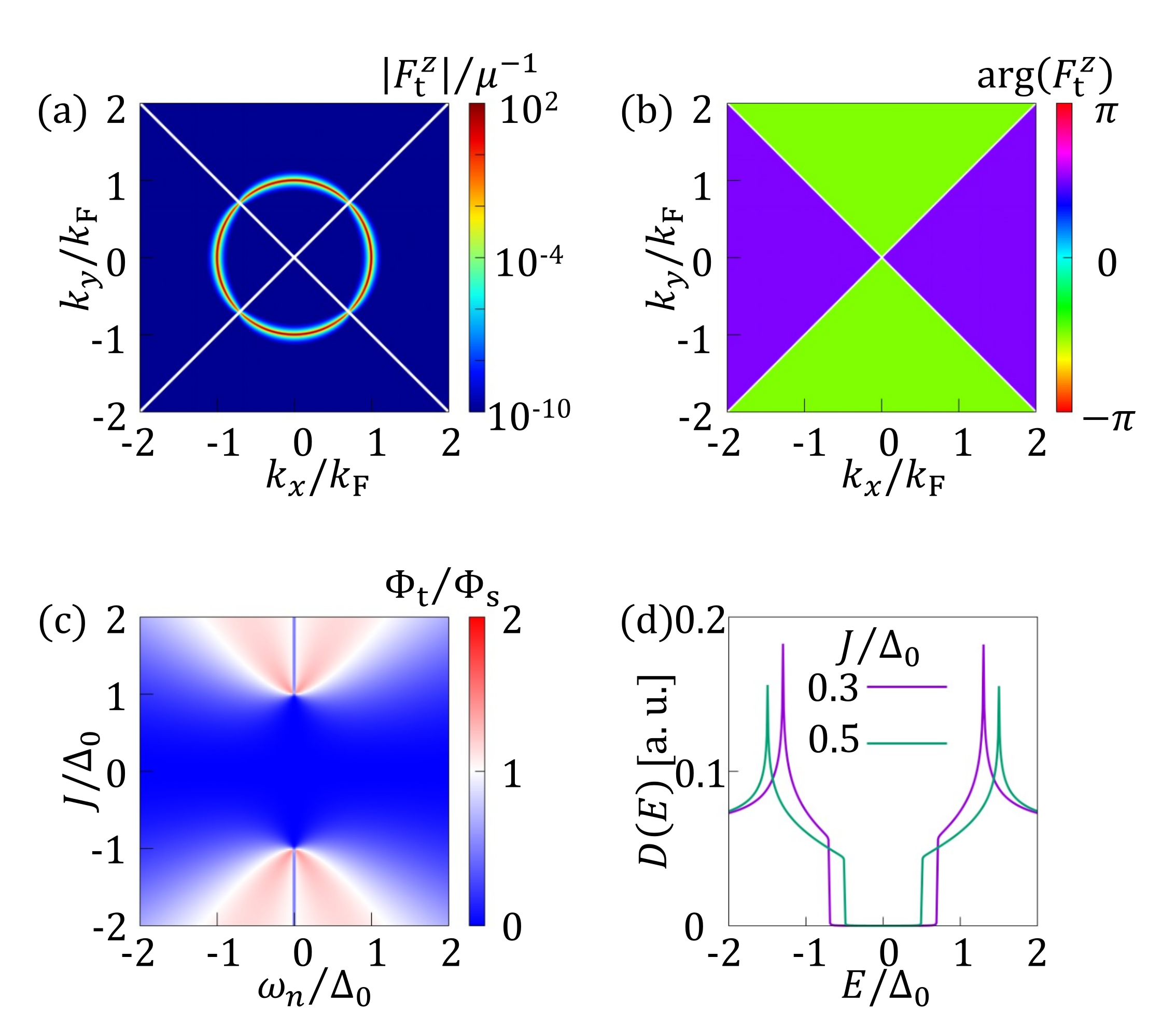}
    \caption{Induced pair amplitudes and DOS in a spin-singlet $s$-wave superconductor with a $d_{x^{2}-y^{2}}$-wave AM. (a) Absolute value of the emergent OTE pair amplitude given by Eq.\,(\ref{Fstsingletdwave})  as a function of momenta.  The diagonal lines indicate the nodes of the OTE pair amplitude, namely, where it becomes zero. 
    (b) Argument of the OTE pair amplitude as a function of momenta, where the {purple and green} colors indicate the argument values of $\pi/2$ and $-\pi/2$, respectively. In (a,b), $\omega_n=0.5\Delta_0$ and $J=0.3\Delta_{0}$. (c) Ratio of the integrated pair potentials   $\Phi_{\mathrm{t}}(i\omega_n)/\Phi_{\mathrm{s}}(i\omega_n)$ as a  function of the exchange energy $J$ and the Matsubara frequency $\omega_n$.  The blue, white, and red colors correspond to  $\Phi_{\mathrm{t}}<\Phi_{\mathrm{s}}$, $\Phi_{\mathrm{t}}=\Phi_{\mathrm{s}}$, and $\Phi_{\mathrm{t}}>\Phi_{\mathrm{s}}$, respectively.   (d) DOS in arbitrary units (a. u.) as a function of energy for distinct values of $J$. Parameters: $\Delta_0=0.01\mu$.}
    \label{Fig2}
\end{figure}

\begin{figure}
    \centering
    \includegraphics[width=0.95\linewidth]{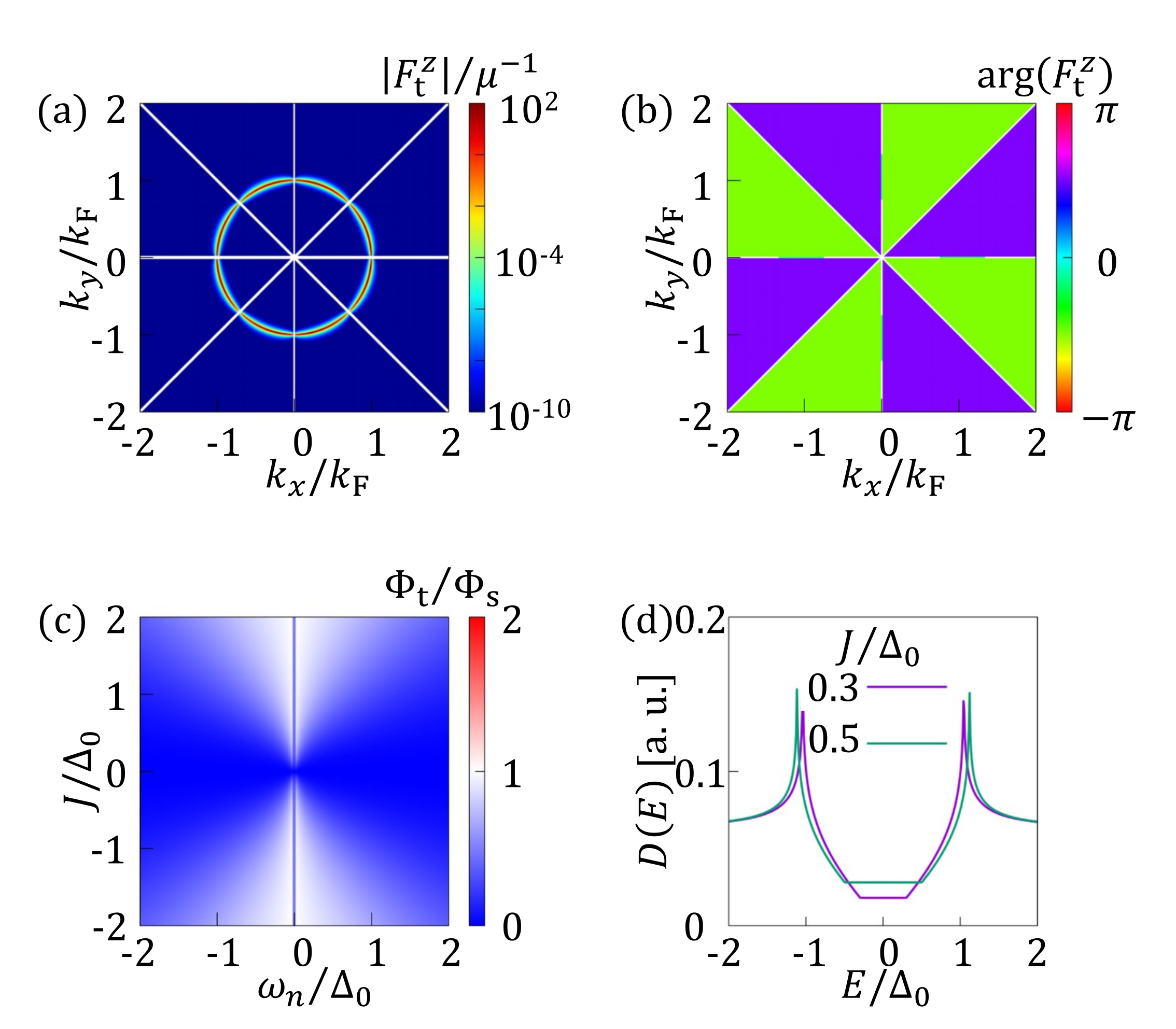}
    \caption{Induced pair amplitudes and DOS in a spin-singlet $d_{xy}$-wave superconductor with a $d_{x^{2}-y^{2}}$-wave AM. (a) Absolute value of the emergent OTE pair amplitude given by Eq.\,(\ref{Fstsingletdwave})  as a function of momenta.  The diagonal  lines indicate the nodes of the pair amplitude. (b) Argument of the OTE pair amplitude as a function of  momenta, where  the purple and green colors indicate $\pi/2$ and $-\pi/2$, respectively.  In (a,b), $\omega_n=0.5\Delta_0$ and $J=0.3\Delta_{0}$.  (c) Ratio of the integrated pair potentials   $\Phi_{\mathrm{t}}(i\omega_n)/\Phi_{\mathrm{s}}(i\omega_n)$ as a  function of the exchange energy $J$ and  Matsubara frequency $\omega_n$.  The blue, white, and red colors indicate  $\Phi_{\mathrm{t}}<\Phi_{\mathrm{s}}$, $\Phi_{\mathrm{t}}=\Phi_{\mathrm{s}}$, and $\Phi_{\mathrm{t}}>\Phi_{\mathrm{s}}$, respectively.  (d) DOS as a function of energy for distinct values of $J$.  Parameters: $\Delta_0=0.01\mu$. }
    \label{Fig3}
\end{figure}

 To visualize the emergence of superconducting correlations given by Eqs.\,(\ref{Fstsingletdwave}), in Fig.\,\ref{Fig2}(a,b) we plot the absolute value and argument of the OTE pair amplitude for a spin-singlet $s$-wave superconductor with a $d_{x^2-y^2}$-wave AM  as a function of momenta \footnote{We note that the data shown in Fig.\,\ref{Fig2}(a,b) is directly generated  from Eqs.\,(\ref{Fstsingletdwave}). Similarly, subsequent figures presented in this work are also generated from the  expressions of the corresponding pair amplitudes. Upon reasonable request, we can share the data presented in the figures of this work.}. Moreover,  in Fig.\,\ref{Fig2}(c) we show the ratio between the integrated squared magnitudes of OTE and ESE pair amplitudes as a function of exchange field $J$ and Matsubara frequency $\omega_n$, while in Fig.\,\ref{Fig2}(d) we show the DOS as a function of energy. The first observation in Fig.\,\ref{Fig2}(a) is that the  OTE pair amplitude develops four nodes along the diagonals in momentum space, which stem from the $d$-wave character coming from the AM and therefore defines its even parity, see white  lines in Fig.\,\ref{Fig2}(a). When inspecting the argument of the OTE pairing [Fig.\,\ref{Fig2}(b)], it develops an alternating opposite sign in four regions as one moves around momentum space but still following an even momentum dependence that reflects its even parity symmetry. While under general conditions, both ESE and OTE pairing coexist [Eq.\,(\ref{Fstsingletdwave})], the OTE pairing can dominate when  $J$ is of the order of $\Delta_{0}$, see red regions in  Fig.\,\ref{Fig2}(c). The effect of the exchange field $J$ is also evident in the DOS, where the superconducting gap shrinks as $J$ increases [Fig.\,\ref{Fig2}(d)],  ultimately leading to a gapless structure and a finite DOS for $J>\Delta_{0}$ \cite{Yokoyama2007}. Since the DOS is given by the normal Green's function and can be separated into the pair amplitude contributions [Eq.\,(\ref{G0F})], a gapless DOS signals an OTE dominant regime in a  spin-singlet $s$-wave superconductors with a $d_{x^{2}-y^{2}}$-wave AM.

Following a similar spirit, in Fig.\,\ref{Fig3} we present the same as in Fig.\,\ref{Fig2} but for a spin-singlet $d_{xy}$-wave superconductor with a $d_{x^2-y^2}$-wave AM. The absolute value of the OTE pair amplitude now unveils eight nodes, showing the $g$-wave symmetry arising due to the interplay between $d$-wave AM and $d$-wave superconductor, see the white lines in Fig.\,\ref{Fig3}(a). Similarly, the argument of the OTE pairing exhibits eight regions with alternating opposite signs, confirming its $g$-wave parity symmetry [Fig.\,\ref{Fig3}(b)]. As noted above, the OTE pairing also coexists with its ESE counterpart and acquire similar values but its dominant behavior is rather challenging [Fig.\,\ref{Fig3}(c)]. In relation to the DOS  shown in Fig.\,\ref{Fig3}(d), it has a V-shape already at $J=0$ due to the $d$-wave nature of the parent superconductor. At finite $J$, the DOS acquires finite constant values for energies below $|J|$ [Fig.\,\ref{Fig3}(d)], a regime where    ESE and OTE pairings are equally important [Fig.\,\ref{Fig3}(c)]. The flat DOS in  Fig.\,\ref{Fig3}(d)  is thus a signal of the emergence of OTE pairing in a spin-singlet $d_{xy}$-wave superconductor with a $d_{x^2-y^2}$-wave AM.

\subsection{Spin-triplet superconductors}\label{subsec3b}
In the case of  superconductors with unitary spin-triplet pair potentials, we distinguish two cases, namely, spin-triplet chiral $p$-wave superconductors and spin-triplet helical $p$-wave superconductors, see  Tab.\,\ref{tab:pair_potentials}. We remind that the spin-triplet chiral $p$-wave superconductors  have   mixed spin-triplet Cooper pairs, while the  spin-triplet helical $p$-wave superconductors are host of equal spin Cooper pairs \cite{QiPRL}.  

For spin-triplet chiral $p$-wave superconductors with  spin-triplet pair potentials having a $\bm{d}(\bm{k})$ vector parallel to the $z$-axis and $d_{0}(\bm{k})=0$ given in Tab.\,\ref{tab:pair_potentials}, we obtain the pair amplitudes given by 
\begin{equation}\label{triplet-1F}
  \begin{split}
  F_{\uparrow\uparrow}(\bm{k},i\omega_n) &= F_{\downarrow\downarrow}(\bm{k},i\omega_n) = 0,\\
  F_{\uparrow\downarrow}(\bm{k},i\omega_n) &= 
  -\frac{d_z(\bm{k})}{P_{\bm{k}}(\omega_n) + 2i\omega_nM^{d}_{\bm{k}}},\\
  F_{\downarrow\uparrow}(\bm{k},i\omega_n) &= 
  -\frac{d_z(\bm{k})}{P_{\bm{k}}(\omega_n) - 2i\omega_nM^{d}_{\bm{k}}}\,,
  \end{split}
\end{equation}
where $P_{\bm{k}}(\omega_n)=\omega_n^2 + \xi_{\bm{k}}^2 - [M^{d}_{\bm{k}}]^2 + |d_z(\bm{k})|^2$ is an even function of momentum and frequency. As for spin-singlet superconductors, we now use Eqs.\,(\ref{triplet-1F}) and Eqs.\,(\ref{Fst}) to identify the spin symmetry of the pair amplitudes. Thus, we obtain the spin-singlet and spin-triplet pair amplitudes given by
\begin{equation}
\label{Ftripletup}
  \begin{split}
  F_\mathrm{s}(\bm{k},i\omega_n)=
  \frac{2i\omega_nd_z(\bm{k}) M^{d}_{\bm{k}}}
  {[P_{\bm{k}}(\omega_n)]^2+4[M_{\bm{k}}^{d}]^2\omega_n^2},\\
  \bm{F}_\mathrm{t}(\bm{k},i\omega_n)=
  -\frac{d_z(\bm{k})P_{\bm{k}}(\omega_n)\hat{\bm{z}}}
  {[P_{\bm{k}}(\omega_n)]^2+4[M^{d}_{\bm{k}}]^2\omega_n^2}\,.
  \end{split}
\end{equation}
Taking  into account that the $d$ vector of the parent superconductor is odd in momentum $d_z(\bm{k})=-d_z(-\bm{k})$ because it is $p$-wave, the spin-singlet pair amplitude $F_\mathrm{s}(\bm{k},i\omega_n)$ in Eq\,(\ref{Ftripletup}) exhibits an OSO symmetry while the spin-triplet amplitude $\bm{F}_\mathrm{t}(\bm{k},i\omega_n)$ has an ETO symmetry.
\begin{figure}
    \centering
    \includegraphics[width=0.95\linewidth]{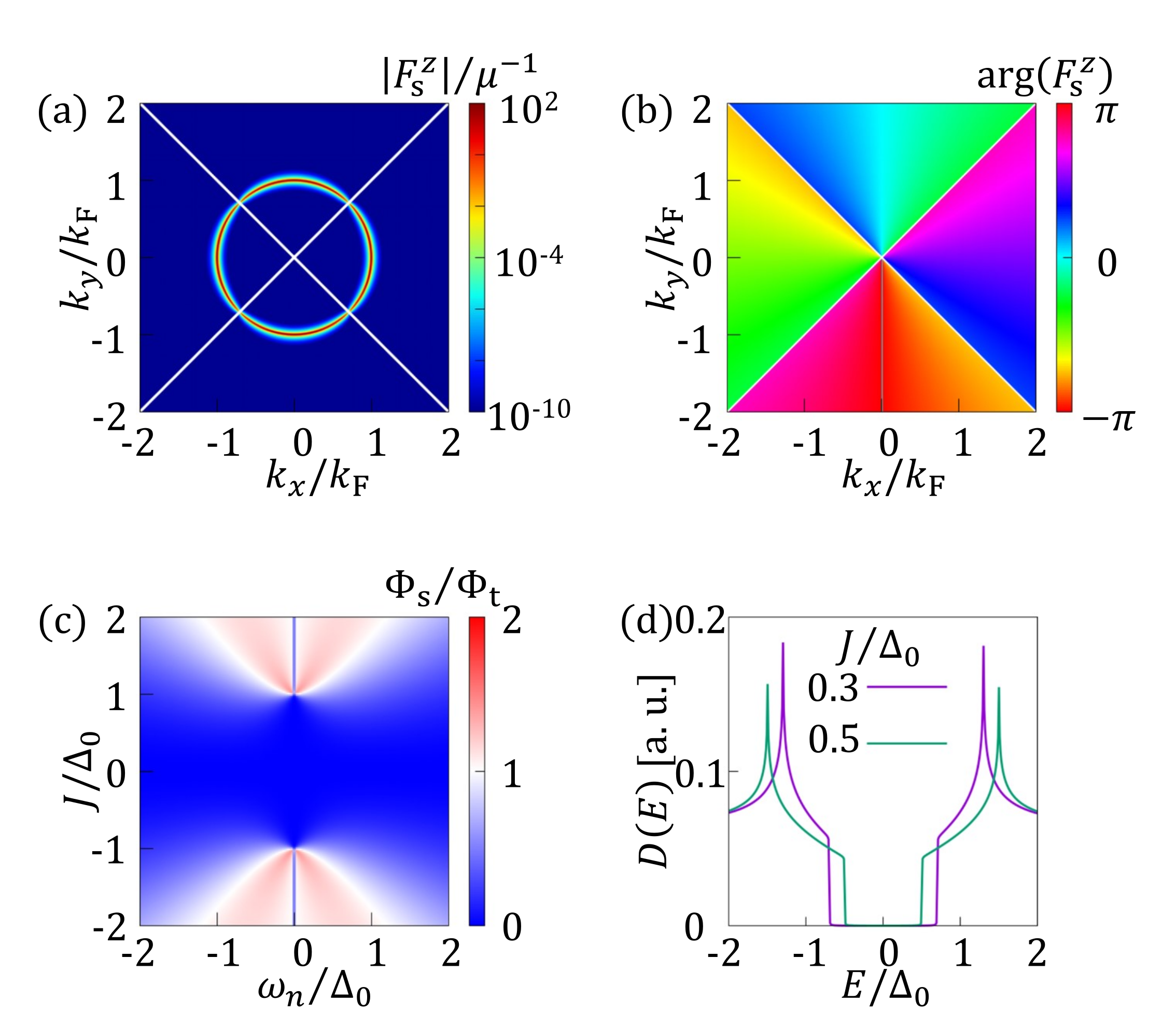}
    \caption{Induced pair amplitudes and DOS in a spin-triplet chiral $p$-wave superconductor with $d_{x^{2}-y^{2}}$-wave AM. (a) The absolute value of the emergent OSO pair amplitude given by Eq.\,(\ref{Ftripletup})  as a function of momenta. The diagonal white lines indicate the nodes of the OSO pair amplitude. (b) Argument of the OSO pair amplitude as a function of momenta,  where the distinct colors show the value of the argument as depicted in the color bar. In (a,b), $\omega_n=0.5\Delta_0$ and $J=0.3\Delta_0$. (c) Ratio of the integrated pair potentials $\Phi_{\mathrm{s}}(i\omega_n)/\Phi_{\mathrm{t}}(i\omega_n)$ as a function of the exchange energy $J$ and the Matsubara frequency $\omega_n$. The blue, white, and red colors depict 
    $\Phi_\mathrm{s}<\Phi_\mathrm{t}$, $\Phi_\mathrm{s}=\Phi_\mathrm{t}$, and $\Phi_\mathrm{s}>\Phi_\mathrm{t}$, respectively. (d) DOS as a function of energy for distinct values of $J$.     Parameters: $\Delta_0=0.01\mu$.}
    \label{Fig4}
\end{figure}
The odd parity symmetry in both pair amplitudes can be understood by noting that the exchange of momentum in the $xy$-plane gives rise to a minus sign in the pair potential [Tab.\,\ref{tab:pair_potentials}], which leads to a global minus sign  when exchanging $\bm{k}$ in $d_{z}(\bm{k})$. It is interesting to note that the parity of the OSO pair amplitude is determined by the interplay of $d_z(\bm{k})$ and $M^{d}_{\bm{k}}$, giving rise to a momentum dependence of $f$-wave nature and hence higher than the parent spin-triplet superconductor. Since the parent superconductor here has a spin-triplet odd-parity pair potential ($d_z(\bm{k})$), the ETO pair symmetry stems from $d_z(\bm{k})$ while the OSO pair correlation is  an emergent effect arising due to the combined effect of $d$-wave altermagnetism and the spin-triplet chiral superconductivity. 
To further understand the above discussion, the absolute value and argument of the OSO pairing is plotted in Figs.\,\ref{Fig4}(a,b) as a function of momenta for a spin-triplet chiral $p$-wave superconductor with a $d_{x^{2}-y^{2}}$-wave AM. We clearly see the four nodes coming from the $d$-wave nature of the AM and the alternating signs [Figs.\,\ref{Fig4}(a,b)], which show the $f$-wave symmetry of the emergent OSO pair amplitude. 
Moreover, the OSO pair amplitude can dominate at $J>\Delta_{0}$, although in general it coexists with the ETO pairing,  see Fig.\,\ref{Fig4}(c). Large values of $J$ also reduce the gap in the DOS and can lead to a finite subgap DOS for $J>\Delta_{0}$,    as seen in Fig.\,\ref{Fig4}(d).   This DOS feature can be then taken as a signal of dominant OSO pairing, see also Eq.\,(\ref{G0F}).

\begin{figure}
    \centering
    \includegraphics[width=0.95\linewidth]{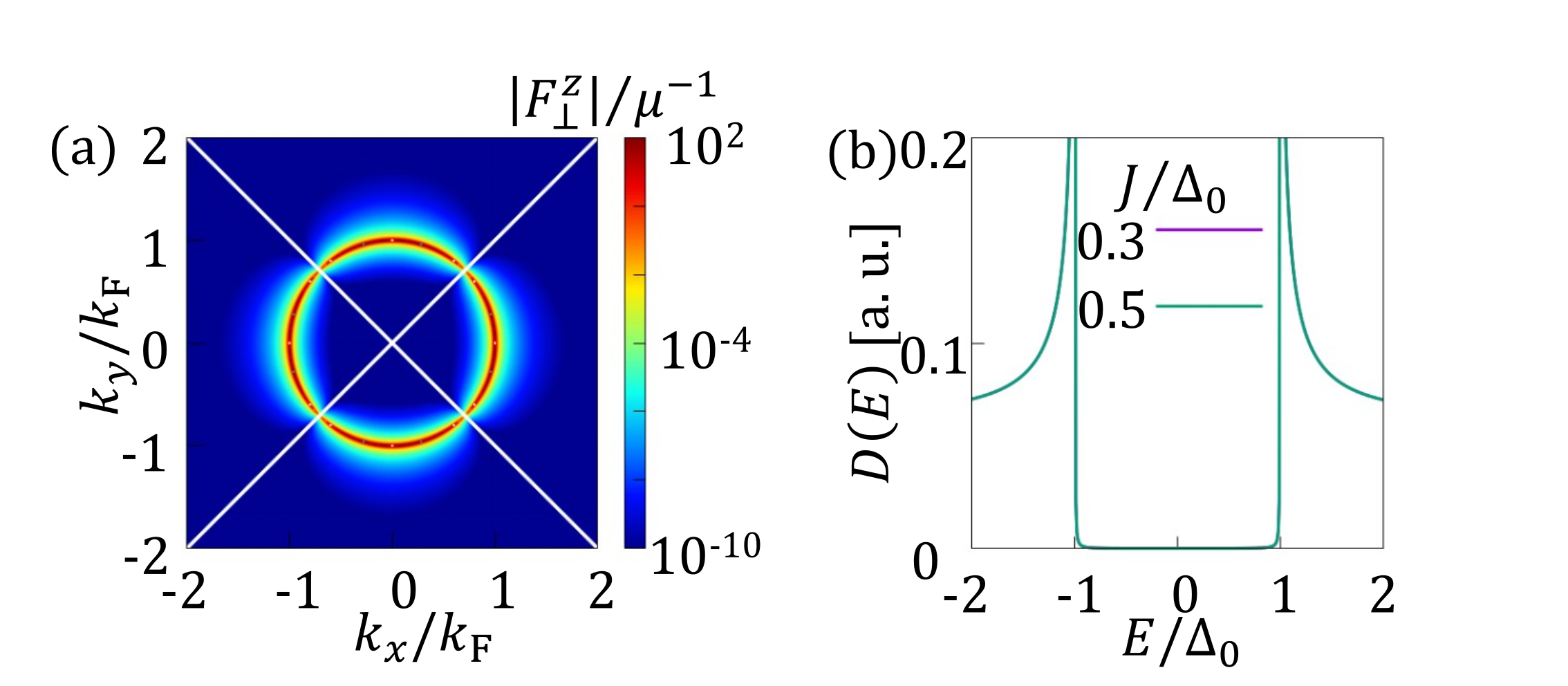}
    \caption{Induced pair amplitudes and DOS in a spin-triplet helical $p$-wave superconductor with $d_{x^{2}-y^{2}}$-wave AM. (a) The absolute value of the emergent pair amplitude $F_{\perp}(\bm{k},i\omega_n)$ given by Eq.\,(\ref{triplet-2Fy}) as a function of momenta at $\omega_{n}=0.5\Delta_{0}$ and $J=0.3\Delta_{0}$.     The diagonal lines indicate the nodes of $F_{\perp}(\bm{k},i\omega_n)$. (b) DOS as a function of energy for distinct values of $J$.      Parameters: $\Delta_0=0.01\mu$.}
    \label{Fig5}
\end{figure}

For   spin-triplet helical $p$-wave superconductors having spin-triplet pair potentials with $\bm{d}(\bm{k})$  vector perpendicular to the $z$-axis   satisfying $\bm{d}^*(\bm{k})\times\bm{d}(\bm{k})=0$ \cite{Frigeri}, and  $d_{0}(\bm{k})=0$, $d_{z}(\bm{k})=0$, we obtain the pair amplitudes as
\begin{equation}
\label{triplet-2F}
  \begin{split}
  F_{\uparrow\downarrow}(\bm{k},i\omega_n) &= F_{\downarrow\uparrow}(\bm{k},i\omega_n) = 0,\\
  F_{\uparrow\uparrow}(\bm{k},i\omega_n) &= 
  \frac{d_x(\bm{k})-id_y(\bm{k})}{\omega_n^2 + (\xi_{\bm{k}} + M^{d}_{\bm{k}})^2 + \bm{d}^*(\bm{k})\cdot\bm{d}(\bm{k})},\\
  F_{\downarrow\downarrow}(\bm{k},i\omega_n) &= 
  \frac{-d_x(\bm{k})-id_y(\bm{k})}{\omega_n^2 + (\xi_{\bm{k}} - M^{d}_{\bm{k}})^2 + \bm{d}^*(\bm{k})\cdot\bm{d}(\bm{k})}\,.
  \end{split}
\end{equation}
Now, we  use Eqs.\,(\ref{triplet-2F}) and Eqs.\,(\ref{Fst}) to decompose the spin symmetry of the pair amplitudes. Then, the  spin-singlet and spin-triplet pair amplitudes are given by
\begin{equation}
\label{triplet-2Fx}
  \begin{split}
  F_\mathrm{s}(\bm{k},i\omega_n) &= 0,\\
  \bm{F}_\mathrm{t}(\bm{k},i\omega_n) &= \bm{F}_{\parallel}(\bm{k},i\omega_n) + \bm{F}_{\perp}(\bm{k},i\omega_n),
  \end{split}
\end{equation}
with
\begin{equation}
\label{triplet-2Fy}
  \begin{split}
  \bm{F}_{\parallel}(\bm{k},i\omega_n) &= 
  -\frac{\bm{d}(\bm{k})[R_{\bm{k}}(\omega_n)]}
  {[R_{\bm{k}}(\omega_n)]^2-4\xi_{\bm{k}}^2[M^{d}_{\bm{k}}]^2},\\
  \bm{F}_{\perp}(\bm{k},i\omega_n) &= 
  \frac{2i\xi_{\bm{k}}M^{d}_{\bm{k}}\hat{\bm{z}}\times\bm{d}(\bm{k})}
  {[R_{\bm{k}}(\omega_n)]^2-4\xi_{\bm{k}}^2[M^{d}_{\bm{k}}]^2}\,,
  \end{split}
\end{equation}
where $R_{\bm{k}}(\omega_n)=\omega_n^2 + \xi_{\bm{k}}^2 + [M^{d}_{\bm{k}}]^2 + \bm{d}^*(\bm{k})\cdot\bm{d}(\bm{k})$ is a function that is even in both momentum and frequency. In Eqs.\,(\ref{triplet-2Fx}), we see that no spin-singlet superconducting pairing is induced, $F_\mathrm{s}(\bm{k},i\omega_n)=0$. This is because the parent helical spin-triplet $p$-wave pair potential represents equal spin Cooper pairs and  altermagnetism cannot mix distinct spins.  Moreover, we find that there is a  spin-triplet pair amplitude $\bm{F}_{\parallel}(\bm{k},i\omega_n)$ originating from the parent spin-triplet pair potential, which has ETO symmetry. Interestingly, there is also a spin-triplet pair amplitude component  $ \bm{F}_{\perp}(\bm{k},i\omega_n)$ that is perpendicular to the parent pair potential and originates entirely due to the interplay of altermagnetism and   spin-triplet helical superconductivity. In fact, while $\bm{F}_{\perp}(\bm{k},i\omega_n)$ requires altermagnetism to be finite via $M^{d}_{\bm{k}}$, it is an odd function under the exchange of momenta which is determined by the nature of the symmetry of the superconductor, namely, $\bm{d}(\bm{k})=-\bm{d}(-\bm{k})$. As a result, the perpendicular pair amplitude $\bm{F}_{\perp}(\bm{k},i\omega_n)$  has ETO symmetry. The odd-parity symmetry of $\bm{F}_{\perp}(\bm{k},i\omega_n)$ is determined by the cubic power in momentum, a quadratic dependence  due to   the $d$-wave AM via $M_{\bm{k}}$ and a linear dependence due to $p$-wave superconductor via $\bm{d}(\bm{k})$,  giving rise to a perpendicular pair amplitude with $f$-wave parity.   The transfer of the $d$-wave symmetry from the AM to the perpendicular spin-triplet pairing $\bm{F}_{\perp}(\bm{k},i\omega_n)$ is seen in Fig.\,\ref{Fig5}(a) via the appearance of four nodes for a spin-triplet helical superconductor with $d_{x^{2}-y^{2}}$-wave altermagnetism.  The DOS corresponding to this system is shown in Fig.\,\ref{Fig5}(b) as a function of energy at distinct values of the exchange field $J$. The DOS very weakly depends on the exchange field, which is not evident by naked eye even for reasonable $J$ values [Fig.\,\ref{Fig5}(b)].  This thus suggests that it is challenging to identify the dominant behaviour of $\bm{F}_{\perp}(\bm{k},i\omega_n)$ from the DOS via Eq.\,(\ref{G0F}).  To close this part, we   stress that the results obtained in this part are also  applicable to other  UMs with higher even-parity angular momentum dependence, as we demonstrate in Appendix \ref{AppendixA}.

\section{Superconductors with $p$-wave magnetism}
\label{section4}
Having discussed superconductors with $d$-wave altermagnetism, in this part we investigate the emergent superconducting correlations in spin-singlet and spin-triplet superconductors with $p$-wave magnetism. The superconductors with $p$-wave magnetism are modelled by Eq.\,(\ref{HBdG}), where the   pair potentials are given by Eq.\,(\ref{DeltaEq}) and Tab.\,\ref{tab:pair_potentials} and the $p$-wave magnets described by Eq.\,(\ref{eq:exchange_UPM}).   Here, we set the direction of the magnetization as $\hat{\bm{n}}=\hat{\bm{z}}$  with $\hat{\bm{z}}$ the unit vector along the $z$-direction. To obtain the emergent pair amplitudes, we follow the discussion presented in Subsection \ref{subsection2a} which involves the calculation of the anomalous Green's functions associated to the BdG Hamiltonian given by Eq.\,(\ref{HBdG}).

\subsection{Spin-singlet superconductors}
In the case of spin-singlet pair potentials characterized by  $d_0(\bm{k})\neq0$ and $\bm{d}(\bm{k})=0$, we find that the components of the anomalous Green's function are given by
\begin{equation}
\label{UPMsingletF}
    \begin{split}
  F_{\uparrow\uparrow}(\bm{k},i\omega_n) &= F_{\downarrow\downarrow}(\bm{k},i\omega_n) = 0,\\
  F_{\uparrow\downarrow}(\bm{k},i\omega_n) &= 
  -\frac{d_0(\bm{k})}{\omega_n^2 + (\xi_{\bm{k}} + M_{\bm{k}})^2 + |d_0(\bm{k})|^2},\\
  F_{\downarrow\uparrow}(\bm{k},i\omega_n) &= 
  \frac{d_0(\bm{k})}{\omega_n^2 + (\xi_{\bm{k}} - M_{\bm{k}})^2 + |d_0(\bm{k})|^2}.
    \end{split}
\end{equation}
Thus, by plugging Eqs.\,(\ref{UPMsingletF}) into Eqs.\,(\ref{Fst}), we decompose the spin symmetry and obtain the spin-singlet and spin-triplet pair amplitudes,
\begin{equation}
\label{Fstsingletpwave}
\begin{split}
  F_\mathrm{s}(\bm{k},i\omega_n)&=
  -\frac{d_0(\bm{k})Q_{\bm{k}}^{p}(\omega_n)}
  {[Q_{\bm{k}}^{p}(\omega_n)]^2-4\xi_{\bm{k}}^2[M_{\bm{k}}^{p}]^2}\,,\\
  \bm{F}_\mathrm{t}(\bm{k},i\omega_n)&=
  \frac{2d_0(\bm{k}) \xi_{\bm{k}}M^{p}_{\bm{k}}\hat{\bm{z}}}
  {[Q_{\bm{k}}^{p}(\omega_n)]^2-4\xi_{\bm{k}}^2[M_{\bm{k}}^{p}]^2}\,.
  \end{split}
\end{equation}
where $Q_{\bm{k}}^{p}(\omega_n)=\omega_n^2+\xi_{\bm{k}}^2 - [M^{p}_{\bm{k}}]^2+|d_0(\bm{k})|^2$ is an even function in both momentum and frequency.

From previous expressions, we directly identify that  $F_\mathrm{s}(\bm{k},i\omega_n)$ has ESE pair symmetry, where the even parity  is dictated by the parent superconductor via $d_0(\bm{k})$ which can be $s$-wave or $d$-wave, see Tab.\,\ref{tab:pair_potentials}. The ESE symmetry of $F_\mathrm{s}(\bm{k},i\omega_n)$ fulfils the antisymmetry condition given by Eq.\,(\ref{Fantisymmetry}). For the spin-triplet pair amplitude $\bm{F}_\mathrm{t}(\bm{k},i\omega_n)$, we find that it is along $z$, has an  even-frequency dependence, and is  odd in momentum $\bm{k}$ due to the $p$-wave magnet via $M^{p}_{\bm{k}}$. The $p$-wave parity of the UPM ($M^{p}_{\bm{k}}=-M^{p}_{-\bm{k}}$) is then transferred to $\bm{F}_\mathrm{t}(\bm{k},i\omega_n)$.  As a result,  $\bm{F}_\mathrm{t}(\bm{k},i\omega_n)$ represents a pair amplitude with ETO  symmetry, which satisfies the antisymmetry condition given by Eq.\,(\ref{Fantisymmetry}).   For a spin-singlet $s$-wave pair potential $d_0(\bm{k})$, the odd parity symmetry of the ETO amplitude $\bm{F}_\mathrm{t}(\bm{k},i\omega_n)$ stems from a  momentum dependence that is linear since $M^{p}_{\bm{k}}$ is linear for a $p$-wave magnet. For a spin-singlet $d$-wave pair potential $d_0(\bm{k})$, the odd parity arises due to a momentum dependence that is cubic, linear due to $M^{p}_{\bm{k}}$ and quadratic due to $d_0(\bm{k})$, implying that the overall odd parity of $\bm{F}_\mathrm{t}(\bm{k},i\omega_n)$ has an $f$-wave nature.

\begin{figure}
    \centering
    \includegraphics[width=0.95\linewidth]{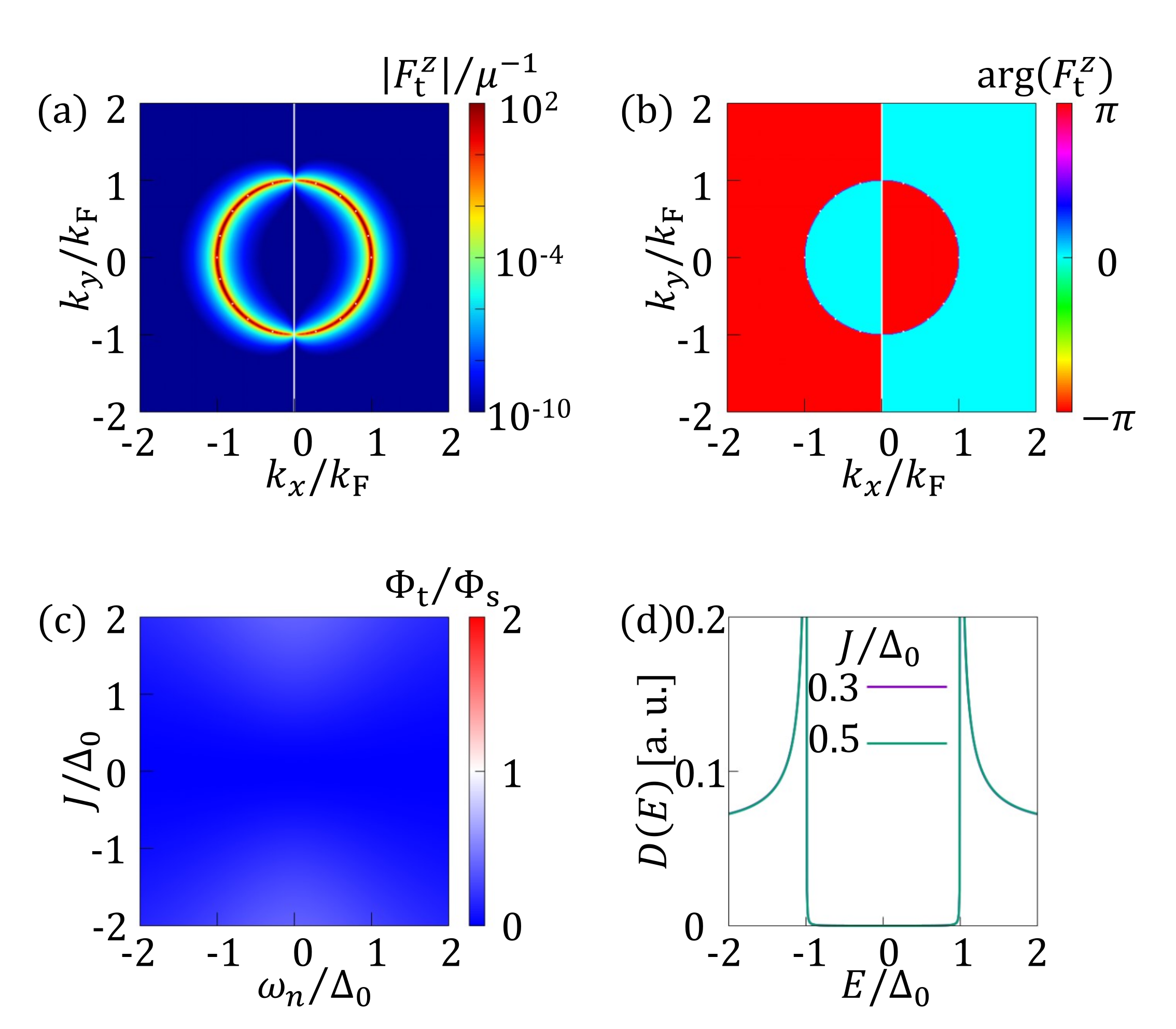}
    \caption{Induced pair amplitudes and DOS in a spin-singlet $s$-wave superconductor with $p_x$-wave magnet. (a) The absolute value of the emergent ETO pair amplitude given by Eq.\,(\ref{Fstsingletpwave})  as a function of momenta. The vertical line indicates the nodes of the ETO pair amplitude.    (b) Argument of the ETO pair amplitude as a function of momenta, where the cyan and red colors indicate the argument values $0$ and $\pm\pi$, respectively. In (a,b), $\omega_n=0.5\Delta_0$ and $J=0.3\Delta_{0}$.    (c) Ratio of the integrated pair potentials $\Phi_{\mathrm{t}}(i\omega_n)/\Phi_{\mathrm{s}}(i\omega_n)$ as a function of the exchange energy $J$ and the Matsubara frequency $\omega_n$. The blue, white, and red colors depict     $\Phi_\mathrm{t}<\Phi_\mathrm{s}$, $\Phi_\mathrm{t}=\Phi_\mathrm{s}$, and $\Phi_\mathrm{t}>\Phi_\mathrm{s}$, respectively. (d) DOS as a function of energy for distinct values of $J$.  Parameters: $\Delta_0=0.01\mu$.}
    \label{Fig6}
\end{figure}

\begin{figure}
    \centering
    \includegraphics[width=0.95\linewidth]{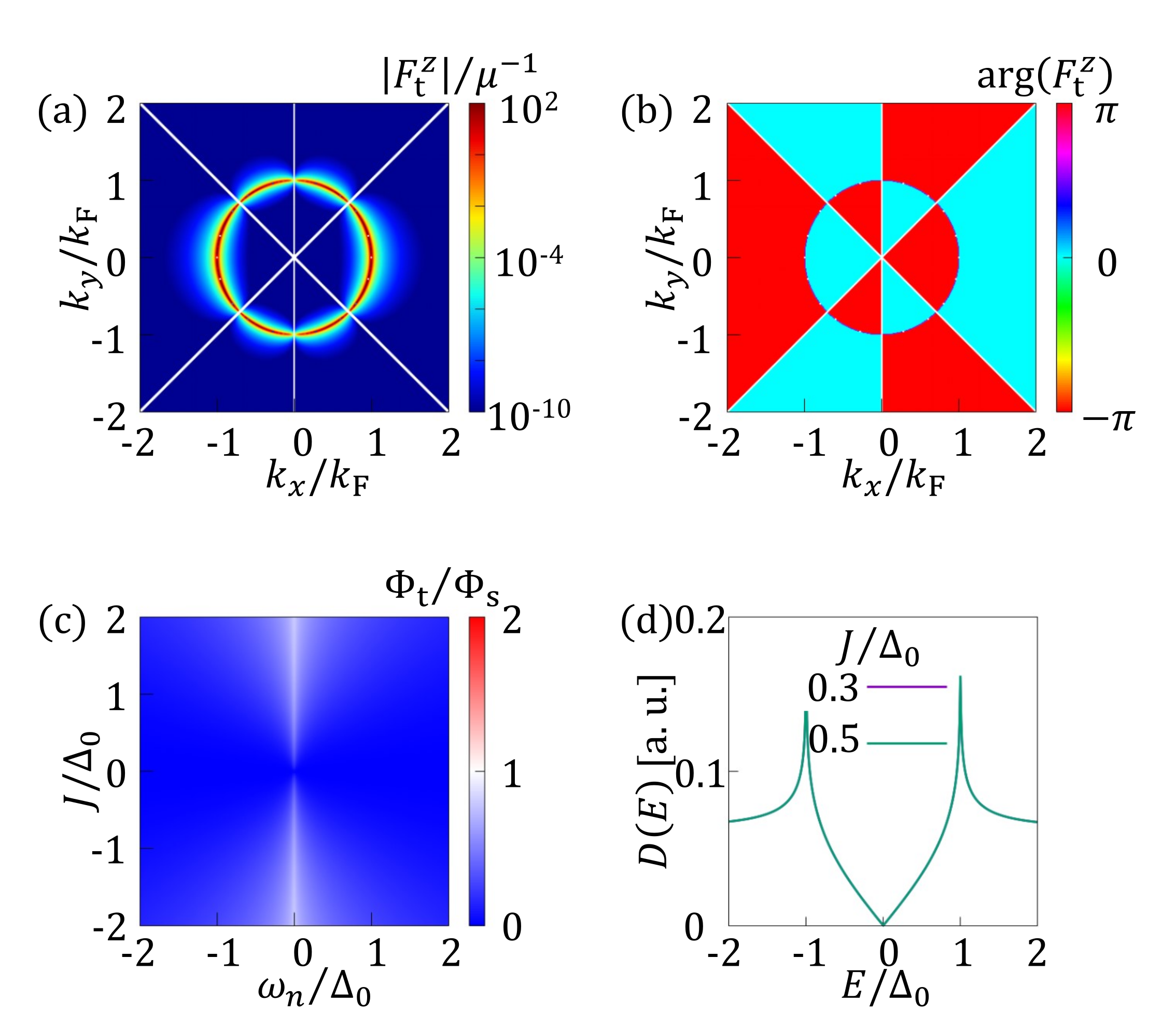} 
    \caption{Induced pair amplitudes and DOS in a spin-singlet $d_{x^2-y^2}$-wave superconductor with $p_x$-wave magnet. (a) The absolute value of the emergent ETO pair amplitude given by Eq.\,(\ref{Fstsingletpwave})  as a function of momenta. The diagonal and vertical lines indicate the nodes of the ETO pair amplitude. (b) Argument of the ETO pair amplitude as a function of momenta, where the cyan and red colors indicate $0$ and $\pm\pi$, respectively. In (a,b), $\omega_n=0.5\Delta_0$ and $J=0.3\Delta_{0}$. 
    (c) Ratio of the integrated pair potentials $\Phi_{\mathrm{t}}(i\omega_n)/\Phi_{\mathrm{s}}(i\omega_n)$ as a function of the exchange energy $J$ and the Matsubara frequency $\omega_n$. The blue, white, and red colors depict 
    $\Phi_\mathrm{t}<\Phi_\mathrm{s}$, $\Phi_\mathrm{t}=\Phi_\mathrm{s}$, and $\Phi_\mathrm{t}>\Phi_\mathrm{s}$, respectively.  (d) DOS as a function of energy for distinct values of $J$. Parameters: $\Delta_0=0.01\mu$.}
    \label{Fig7}
\end{figure}

As an example of the above discussion, in Fig.\,\ref{Fig6}(a,b) we show the absolute value and argument of the emerging ETO pairing $\bm{F}_\mathrm{t}(\bm{k},i\omega_n)$ for a spin-singlet $s$-wave superconductor with a $p_{x}$-wave magnet as a function of momenta. The ETO amplitude exhibits two nodes at $k_{x}=0$ [Fig.\,\ref{Fig6}(a)] and its argument acquires values of $0,\pi$ that reflect the oddness in momentum [Fig.\,\ref{Fig6}(b)]. The circle formed in Fig.\,\ref{Fig6}(b) around zero momenta with changing sign has a radius defined by $k_{\rm F}$ and comes from $\xi_{\bm{k}}$ in the numerator of $\bm{F}_\mathrm{t}(\bm{k},i\omega_n)$ in Eq.\,(\ref{Fstsingletpwave}). The size of the induced ETO pairing hardly overcomes the ESE pairing due to the parent superconductor [Fig.\,\ref{Fig6}(c)], which makes it difficult to be distinguishable in the DOS at reasonable values of $J$ [Fig.\,\ref{Fig6}(d)]. When the superconductor is $d_{x^2-y^2}$-wave and the magnet is $p_{x}$-wave, the ETO pairing develops six nodes marked by white lines in Fig.\,\ref{Fig7}(a) unveiling its $f$-wave symmetry, out of which four stem from the $d$-wave superconductor and two from the $p$-wave magnet. The argument of ETO here picks up $0,\pi$ that show the odd-parity symmetry [Fig.\,\ref{Fig7}(b)]. Moreover, ETO can be equally large as ESE even at weak $J$ [Fig.\,\ref{Fig7}(c)], implying that both contribute to the V-shaped DOS, as seen in Fig.\,\ref{Fig7}(d); see also Eq.\,(\ref{G0F}). The DOS profile originates from the $d$-wave nature of the superconductor and remains largely unchanged for reasonable values of the strength of $J$ for a $p$-wave magnet. The DOS insensitivity to variations of $J$ can be  understood by noting that the effect of $J$ enters via the normal dispersion $\xi_{\bm{k}\sigma}$. In fact,  $\xi_{\bm{k}\sigma}$ can be rewritten as
$ \xi_{\bm{k}\sigma} = [\hbar^2/(2m)]
    \{
     (k_x+\mathrm{sgn}(\sigma)[mJ/(\hbar^2 k_F)]
\cos\alpha)^2+
     (k_y+\mathrm{sgn}(\sigma)[mJ/(\hbar^2 k_F)] 
    \sin\alpha)^2
    \}
    -(\mu+[J^{2}/(4\mu)])
$, where $\mathrm{sgn}(\sigma)$ by $\mathrm{sgn}(\uparrow)=+1$ and $\mathrm{sgn}(\downarrow)=-1$. Hence, $J$ effectively shifts the wavevector's origin and changes the chemical potential's effective value.  Thus, after integrating the spectral function in $\bm{k}$-space, the effect of $\bm{k}$-shift due to $J$ is canceled, which results in the DOS shown in  Fig.\,\ref{Fig7}(d).

{ We further point out that, when combining $d_{xy}$-wave superconductivity with $p_{y}$-wave magnetism, the ETO pair amplitude is an odd function of $k_x$ and an even function of $k_y$, thus having an odd-parity symmetry as when combining $d_{xy}$-wave superconductivity with $p_{x}$-wave magnetism discussed above. In this regard, it is worth noting that the emergence of ETO pair correlations have been predicted before in the bulk of a spin-singlet $d_{xy}$-wave superconductor with a persistent spin helix that has a  spin-splitting similar to the $p_{y}$-wave UPMs studied here, see Refs.\,\cite{LeePRB2021,ikegaya2021strong}. Our findings thus show their applicability to other systems beyond spin-singlet superconductors with unconventional $p$-wave magnets.}

\subsection{Spin-triplet superconductors}
We now obtain the emergent pair amplitudes for spin-triplet superconductors with $p$-wave magnetism. We consider spin-triplet chiral and spin-triplet helical $p$-wave superconductors with pair potentials given in Tab.\,\ref{tab:pair_potentials}, which characterize mixed-spin triplet and equal-spin triplet Cooper pairs, respectively.

\begin{figure}
    \centering
    \includegraphics[width=0.95\linewidth]{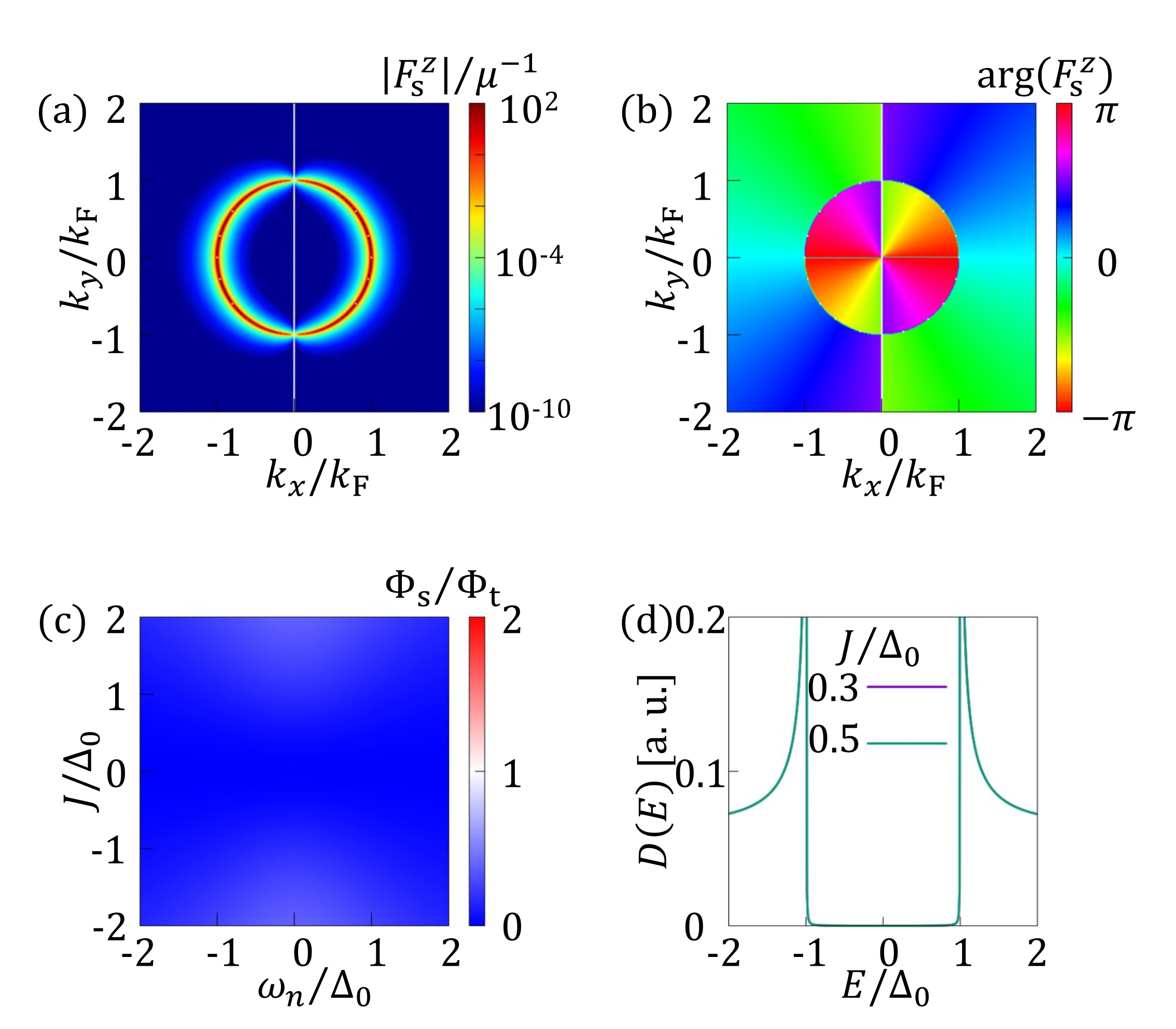}
    \caption{Induced pair amplitudes and DOS in a spin-triplet chiral $p$-wave superconductor with $p_x$-wave magnet. (a) The absolute value of the emergent ESE pair amplitude given by Eq.\,(\ref{Fsttriplet-1pwave})  as a function of momenta. The vertical lines indicate the nodes of the ESE pair amplitude. (b) Argument of the ESE pair amplitude as a function of momenta, where the distinct colors show the value of the argument as depicted in the color bar. In (a,b), $\omega_n=0.5\Delta_0$ and $J=0.3\Delta_0$. (c) Ratio of the integrated pair potentials $\Phi_{\mathrm{s}}(i\omega_n)/\Phi_{\mathrm{t}}(i\omega_n)$ as a function of the exchange energy $J$ and the Matsubara frequency $\omega_n$. The blue, white, and red colors depict 
    $\Phi_\mathrm{s}<\Phi_\mathrm{t}$, $\Phi_\mathrm{s}=\Phi_\mathrm{t}$, and $\Phi_\mathrm{s}>\Phi_\mathrm{t}$, respectively.     (d) DOS as a function of energy for distinct values of $J$.    Parameters: $\Delta_0=0.01\mu$.}
    \label{Fig8}
\end{figure}

For spin-triplet chiral $p$-wave superconductors, we consider the pair potential determined by $\bm{d}(\bm{k})\parallel\hat{\bm{z}}$ and $d_0(\bm{k})=0$. Hence, the components of the anomalous Green's functions are given by
\begin{equation}\label{UPMtriplet-1F}
    \begin{split}
  F_{\uparrow\uparrow}(\bm{k},i\omega_n) &= F_{\downarrow\downarrow}(\bm{k},i\omega_n) = 0,\\
  F_{\uparrow\downarrow}(\bm{k},i\omega_n) &= 
  -\frac{d_z(\bm{k})}{\omega_n^2 + (\xi_{\bm{k}} + M^{p}_{\bm{k}})^2 + |d_z(\bm{k})|^2},\\
  F_{\downarrow\uparrow}(\bm{k},i\omega_n) &= 
  -\frac{d_z(\bm{k})}{\omega_n^2 + (\xi_{\bm{k}} - M^{p}_{\bm{k}})^2 + |d_z(\bm{k})|^2}.
    \end{split}
\end{equation}
We can now   use  Eqs.\,(\ref{UPMtriplet-1F}) and Eqs.\,(\ref{Fst}) to obtain the spin-singlet and spin-triplet pair amplitudes, which read
\begin{equation}
\label{Fsttriplet-1pwave}
\begin{split}
  F_\mathrm{s}(\bm{k},i\omega_n)&=
  \frac{2 \xi_{\bm{k}}M^{p}_{\bm{k}}d_z(\bm{k})}
  {[P^{p}_{\bm{k}}(\omega_n)]^2-4\xi_{\bm{k}}^2[M^{p}_{\bm{k}}]^2}\,.
\\
  \bm{F}_\mathrm{t}(\bm{k},i\omega_n)&=
  -\frac{P^{p}_{\bm{k}}(\omega_n)d_z(\bm{k})\hat{\bm{z}}}
  {[P^{p}_{\bm{k}}(\omega_n)]^2-4\xi_{\bm{k}}^2[M^{p}_{\bm{k}}]^2}\,,
  \end{split}
\end{equation}
where $P^{p}_{\bm{k}}(\omega_n)=\omega_n^2+\xi_{\bm{k}}^2 - [M^{p}_{\bm{k}}]^2+|d_z(\bm{k})|^2$ is an even function in both momentum and frequency.
We realize that the spin-singlet pair amplitude $F_\mathrm{s}(\bm{k},i\omega_n)$   is even in momentum, which, interestingly, occurs due to the simultaneous effect of chiral $p$-wave superconductivity and $p$-wave magnetism via $d_z(\bm{k})$ and $M_{\bm{k}}^{p}$, respectively. In fact, both effects are odd in momentum, namely, $d_z(\bm{k})=-d_z(-\bm{k})$ and $M_{\bm{k}}^{p}=-M_{-\bm{k}}^{p}$, hence resulting in an even in momentum dependence. Hence, $F_\mathrm{s}(\bm{k},i\omega_n)$ has an ESE symmetry, consistent with the antisymmetry condition dictated by Eq.\,(\ref{Fantisymmetry}). The even parity of $F_\mathrm{s}(\bm{k},i\omega_n)$ is given by a quadratic dependence on momentum, which can then be interpreted as a sort of $d$-wave symmetry.
When it comes to the spin-triplet pair amplitude $\bm{F}_\mathrm{t}(\bm{k},i\omega_n)$, it is linearly proportional to $d_z(\bm{k})$, which then determines its odd-parity symmetry directly coming from the chiral $p$-wave nature of the parent superconductor. These dependences classify $\bm{F}_\mathrm{t}(\bm{k},i\omega_n)$ as an ETO pair amplitude.  To visualize the behaviour of the emerging ESE pairing discussed here, in Fig.\,\ref{Fig8}(a,b) we show its absolute value and its argument as a function of momenta. As expected, it has two nodes at $k_{x}=0$ due to the $p$-wave magnet and its argument acquires values that show the evenness in momentum, while developing an inner circle defined by $\xi_{\bm{k}}$, see Fig.\,\ref{Fig8}(b). The size of ESE is rather small in comparison to ETO, albeit large $J$ can induce sizeable values but still smaller [Fig.\,\ref{Fig8}(c)]. This implies that identifying the contribution from the ESE pairing in the DOS is challenging [Fig.\,\ref{Fig8}(d)]. 
 
\begin{figure}
    \centering
    \includegraphics[width=0.95\linewidth]{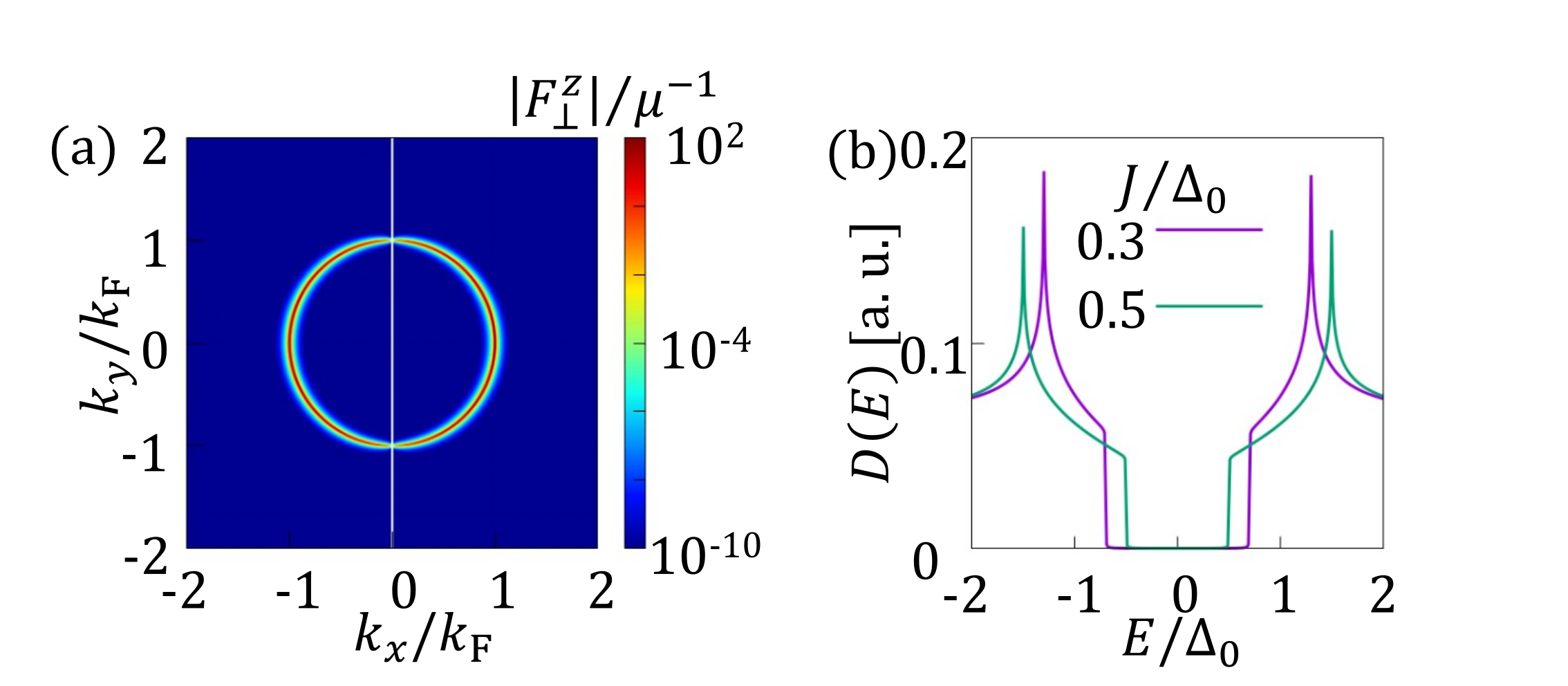}
    \caption{Induced pair amplitudes and DOS in a spin-triplet helical $p$-wave superconductor with $p_{x}$-wave magnet. (a) The absolute value of the emergent OTE pair amplitude given by Eq.\,(\ref{tripletpwaveMag2}) as a function of momenta at $\omega_{n}=0.5\Delta_{0}$ and $J=0.3\Delta_0$.     The vertical line indicates the nodes of the  OTE pair amplitude. (b) DOS as a function of energy for distinct values of $J$.   Parameters: $\Delta_0=0.01\mu$.}
    \label{Fig9}
\end{figure}

When it comes to spin-triplet helical $p$-wave superconductors with  pair potentials having a $\bm{d}(\bm{k})$ vector perpendicular to the $z$-axis as $\bm{d}^*(\bm{k})\times\bm{d}(\bm{k})=0$ and $d_0(\bm{k})=0$, we find 
\begin{equation}\label{UPMtriplet-2F}
  \begin{split}
  F_{\uparrow\downarrow}(\bm{k},i\omega_n) &= F_{\downarrow\uparrow}(\bm{k},i\omega_n) = 0,\\
  F_{\uparrow\uparrow}(\bm{k},i\omega_n) &= 
  \frac{d_x(\bm{k})-id_y(\bm{k})}{R^{p}_{\bm{k}}(\omega_n) + i2M^{p}_{\bm{k}}\omega_n},\\
  F_{\downarrow\downarrow}(\bm{k},i\omega_n) &= 
  \frac{-d_x(\bm{k})-id_y(\bm{k})}{R^{p}_{\bm{k}}(\omega_n)- i2M^{p}_{\bm{k}}\omega_n}\,.
  \end{split}
\end{equation}
where $R^{p}_{\bm{k}}(\omega_n)=\omega_n^2 + \xi_{\bm{k}}^2 - [M^{p}_{\bm{k}}]^2 + \bm{d}^*(\bm{k})\cdot\bm{d}(\bm{k})$
Now, combining Eqs.\,(\ref{UPMtriplet-2F}) and Eqs.\,(\ref{Fst}), we obtain the spin-singlet and spin-triplet  pair amplitudes,
\begin{equation}
\label{tripletpwaveMag1}
  \begin{split}
  F_\mathrm{s}(\bm{k},i\omega_n) &= 0,\\
  \bm{F}_\mathrm{t}(\bm{k},i\omega_n) &= \bm{F}_{\parallel}(\bm{k},i\omega_n) + \bm{F}_{\perp}(\bm{k},i\omega_n),
  \end{split}
\end{equation}
where
\begin{equation}
\label{tripletpwaveMag2}
  \begin{split}
  \bm{F}_{\parallel}(\bm{k},i\omega_n) &= 
  -\frac{\bm{d}(\bm{k})R^{p}_{\bm{k}}(\omega_n)}
  {[R^{p}_{\bm{k}}(\omega_n)]^2+4\omega_{n}^{2}[M^{p}_{\bm{k}}]^{2}},\\
  \bm{F}_{\perp}(\bm{k},i\omega_n) &= 
  -\frac{2 \omega_{\bm{k}}M^{p}_{\bm{k}}\hat{\bm{z}}\times\bm{d}(\bm{k})}
  {[R^{p}_{\bm{k}}(\omega_n)]^2+4\omega_{n}^{2}[M^{p}_{\bm{k}}]^{2}}\,.
  \end{split}
\end{equation}
Eq.\,(\ref{tripletpwaveMag1}) shows that no spin-singlet superconducting correlations are induced by combining $p$-wave magnetism and spin-triplet helical superconductors. Moreover, Eq.\,(\ref{tripletpwaveMag1})  demonstrates that the spin-triplet pair amplitude develops two orthogonal components, $\bm{F}_{\parallel,\perp}(\bm{k},i\omega_n)$. There emerges a pair amplitude  $\bm{F}_{\parallel}(\bm{k},i\omega_n)$ that is parallel to the  $\bm{d}(\bm{k})$  vector of the parent superconductor, with odd parity due to $\bm{d}(\bm{k})=-\bm{d}(-\bm{k})$, thus classifying $\bm{F}_{\parallel}(\bm{k},i\omega_n)$  as an ETO pair symmetry class coming from the parent superconductor and in line with Eq.\,(\ref{Fantisymmetry}). 
In the spin-triplet pair amplitude we also obtain $\bm{F}_{\perp}(\bm{k},i\omega_n)$ which is perpendicular to the  $\bm{d}(\bm{k})$  vector of the parent superconductor.  This spin-triplet pair amplitude is odd in frequency and has even parity, and hence belongs to class OTE. The even parity pair symmetry is a result of the combined effect of the $p$-wave magnet via  $M^{p}_{\bm{k}}$ and the the parent superconductor via $\bm{d}(\bm{k})$. While both $M^{p}_{\bm{k}}$ and $\bm{d}(\bm{k})$  induce a linear in momentum dependence, the resulting momentum dependence of $\bm{F}_{\perp}(\bm{k},i\omega_n)$ is a quadratic dispersion and hence behave as a $d$-wave symmetry.  The emerging OTE pairing $\bm{F}_{\perp}(\bm{k},i\omega_n)$  is shown in Fig.\,\ref{Fig9}(a) as a function of momenta, where we note its vanishing value at $k_{x}$ due to the nature of the  $p$-wave magnet. In this case, the effect of $J$ in the DOS is that it reduces the energy gap [Fig.\,\ref{Fig9}(b)], which can fill up for $J>\Delta_{0}$ with a sizeable contribution due to the emerging perpendicular OTE pairing.  The discussion presented here can be further generalized to UMs with higher odd-parity momentum dependence, as discussed in Appendix \ref{AppendixA}.

\begin{table*}
\centering
  \caption{
List of emergent pair symmetry classes when combining superconductivity with unconventional magnetism. Top row and leftmost column indicate the type of superconductor and unconventional magnet, respectively. From the second column and second row, the symmetries of the induced pair amplitudes are indicated. For instance, the combination of spin-singlet $s$-wave superconductivity with $d$-wave AM results in two pair symmetry classes which are $s$-wave ESE and $d$-wave OTE, with a $d$-vector parallel to the $z$-axis. The symbols ``$s, p, d, f,\dots$-wave''   indicate that the pair amplitudes can be written as homogeneous polynomials of momenta ($k_x$ and $k_y$) of degree $0,1,2,3,\dots$, respectively.
    }
\begin{tabular}{||c||c| c| c| c||} 
 \hline
{\bf  Magnet / SC} & {\bf singlet $s$-wave} & {\bf singlet $d$-wave} & {\bf triplet chiral $p$-wave}&{\bf triplet helical $p$-wave} \\ [0.5ex] 
 \hline\hline
 {\bf $d$-wave altermagnet}&
 \begin{tabular}{c}
$s$-wave ESE \\ 
$d$-wave OTE($\parallel\hat{\bm{z}}$) 
\end{tabular} &
 \begin{tabular}{c}
$d$-wave ESE \\ 
$g$-wave OTE($\parallel\hat{\bm{z}}$) 
\end{tabular}  & 
 \begin{tabular}{c}
$p$-wave ETO($\parallel\bm{d}$)  \\ 
$f$-wave OSO 
\end{tabular}  & 
 \begin{tabular}{c}
$p$-wave ETO($\parallel\bm{d}$) \\ 
$f$-wave ETO($\parallel\hat{\bm{z}}\times\bm{d}$) 
\end{tabular}  \\ 
 \hline
  {\bf $p$-wave magnet}& 
  \begin{tabular}{c}
$s$-wave ESE \\ 
$p$-wave ETO($\parallel\hat{\bm{z}}$)  
\end{tabular}  & 
 \begin{tabular}{c}
$d$-wave ESE \\ 
$f$-wave ETO($\parallel\hat{\bm{z}}$)  
\end{tabular}  &  
\begin{tabular}{c}
$p$-wave ETO($\parallel\bm{d}$) \\ 
$d$-wave ESE 
\end{tabular}  & 
 \begin{tabular}{c}
$p$-wave ETO($\parallel\bm{d}$) \\ 
$d$-wave OTE($\parallel\hat{\bm{z}}\times\bm{d}$) 
\end{tabular}  \\ [1ex] 
 \hline
\end{tabular}
\label{tab:pair_amp_symmetry}
 \end{table*}
\section{Conclusions}
\label{section5}
In conclusion, we have investigated the emergence of superconducting correlations in unconventional magnets with spin-singlet and spin-triplet superconductivity. We have demonstrated that unconventional magnets induce a spin-singlet to mixed spin-triplet conversion in spin-singlet and spin-triplet chiral $p$-wave superconductors, while such a conversion is absent in spin-triplet helical $p$-wave superconductors. Moreover, we have shown that unconventional magnets transfer their parity symmetry to the emerging superconducting correlations, which makes them to have a higher degree in their momentum dependence. In the case of conventional spin-singlet $s$-wave superconductors, we found that $d$-wave altermagnetism and $p$-wave magnetism can induce, respectively, odd-frequency mixed spin-triplet $d$-wave and even-frequency mixed spin-triplet $p$-wave superconducting correlations. For spin-singlet $d$-wave superconductors, we obtained that $d$-wave altermagnets can host odd-frequency mixed spin-triplet $g$-wave pairing, while an even-frequency mixed spin-triplet $f$-wave pairing emerges in $p$-wave magnets. When combining spin-triplet chiral $p$-wave superconductors with $d$-wave altermagnetism and $p$-wave magnetism, we showed that odd-frequency spin-singlet $f$-wave and even-frequency spin-singlet $d$-wave superconducting correlations are induced, respectively. Finally, for spin-triplet helical $p$-wave superconductors, we found that even-frequency equal spin  $f$-wave and odd-frequency equal spin $d$-wave pair amplitudes form in $d$-wave altermagnets and $p$-wave magnets, respectively.  Table~\ref{tab:pair_amp_symmetry} summarizes the emergent pair symmetries due to the combined effect of superconductivity and unconventional magnetism. Our findings offer a comprehensive discussion of the possible emergent superconducting correlations in unconventional magnets.

\textit{Note added.} Recently, we became aware of Refs.\cite{chakraborty20242,sukhachov2025coexis,chatterjee2025interplay}, which  partially overlap with some of the topics studied in our present work.


\section{Acknowledgements}
K. M. thanks S.\ Ikegaya for valuable discussions. Y. F. and K. Y. acknowledge financial support from the Sumitomo Foundation.  B. L. acknowledges financial support from the National Natural Science Foundation of China (project 12474049).   Y. T. acknowledges financial support from JSPS with Grants-in-Aid for Scientific research  (KAKENHI Grants No. 23K17668  and 24K00583). J. C. acknowledges financial support from the Swedish Research Council  (Vetenskapsr\aa det Grant No.~2021-04121) and the Carl Trygger’s Foundation (Grant No. 22: 2093).


\setcounter{figure}{0}
\renewcommand{\thefigure}{\Alph{section}\arabic{figure}}
\appendix
\section{Superconductors with $f$-, $g$-, and $i$-wave magnets}
\label{AppendixA}
In this appendix, we  highlight that the obtained expressions for the emergent superconducting correlations $F_\mathrm{s}(i\omega_n,\bm{k})$ and $\bm{F}_\mathrm{t}(i\omega_n,\bm{k})$ in sections~\ref{section2} and \ref{section3} are also applicable to superconductors with more exotic unconventional magnetism having the magnetization direction as $\hat{\bm{n}}=\hat{\bm{z}}$ such as due to magnets with $f$-, $g$-, and  $i$-wave   symmetry.  This is because the effective exchange field  of  $g$-, and  $i$-wave magnets exhibits the same even parity symmetry as the $d$-wave AMs studied in this work, $M_{\bm{k}}=M_{-\bm{k}}$; hence,   the results of Section \ref{section3} on AMs are directly applicable to superconductors with  $g$-, and  $i$-wave magnets. Similarly, $f$-wave magnets exhibit an  effective exchange field  that is odd in momentum, $M_{\bm{k}}=-M_{-\bm{k}}$; this implies that the results of Section \ref{section4} on $p$-wave magnets are valid for superconductors with $f$-wave magnets. Of course that the parity of the induced pairing also develops a higher momentum dependence as discussed in Sections \ref{section3} and \ref{section4}.
 
\begin{figure}
    \centering
    \includegraphics[width=0.95\linewidth]{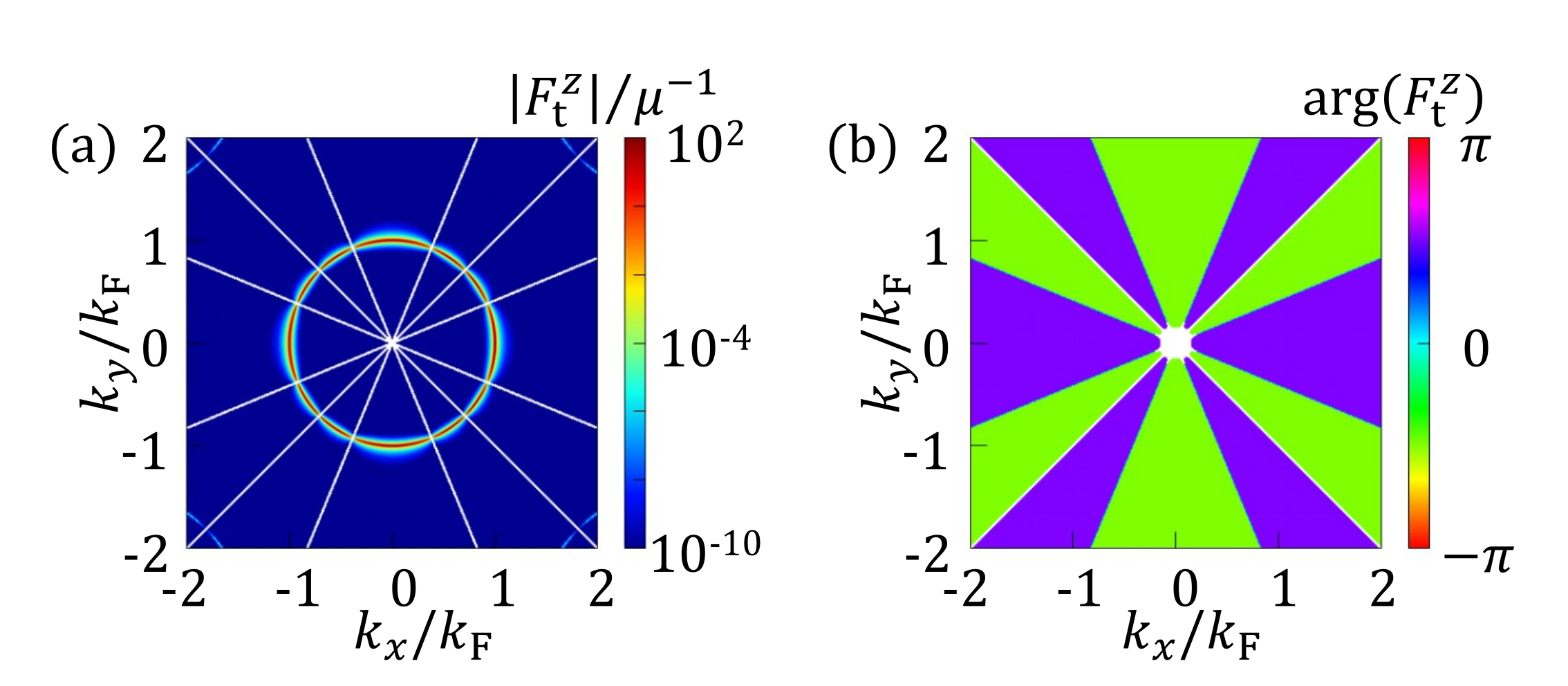}
    \caption{Induced pair amplitudes and DOS in a spin-singlet $d_{x^2-y^2}$-wave superconductor with a $g$-wave AM. (a) Absolute value of the emergent OTE pair amplitude given by Eq.\,(\ref{Fstsingletdwave})  as a function of momenta, where the white lines indicate the nodes of the OTE pair amplitude. (b) Argument of the OTE pair amplitude as a function of momenta, where the purple and green colors indicate the values of $\pi/2$ and $-\pi/2$, respectively. Parameters: $\Delta_0=0.01\mu$, $\omega_n=0.5\Delta_0$, and $J=0.3\Delta_{0}$, $\alpha=0$.}
    \label{Fig10}
\end{figure}

As an example, we consider the case of a spin-singlet $d_{x^{2}-y^{2}}$-wave superconductor with a $g$-wave AM. A generic expression for the exchange field of a $g$-wave AM can be written as 
 \begin{equation}
 \label{gAM}
 \begin{split}
 M_{\bm{k}}^{g} &= \frac{J}{k_{\mathrm{F}}^4}
 \Big[(k_x^4-6k_x^2k_y^2+k_y^4)\cos{4\alpha}\\
 &+4k_xk_y(k_x^2-k_y^2)\sin{4\alpha}\Big],
 \end{split}
 \end{equation}
 where $\alpha$  represents the angle between the $x$-axis and the lobe of AM. Note that for $\alpha=\pi/8$, Eq.\,(\ref{gAM}) reduces to the expression for the $g$-wave AM given below Eq.\,(\ref{eq:exchange_UPM}) up to a factor of $4$. Since this type of AM has an even parity, namely, $M_{\bm{k}}^{g}=M_{-\bm{k}}^{g}$, the emergent pair amplitudes are in this case given by Eqs.\,(\ref{Fstsingletdwave}). Hence, the combination of the $g$-wave nature of the AM and the $d$-wave nature of the superconductor originates an emergent OTE pairing with sextic momentum dependence (degree 6). To visualize this exotic emergent OTE pairing, in Fig.\,\ref{Fig10}(a,b) we plot its magnitude and argument as a function of momenta at $\alpha=0$.  It is interesting to see that this induced OTE pairing has 12 nodes [Fig.\,\ref{Fig10}(a)] and its argument develops an alternating opposite sign when going around momentum [Fig.\,\ref{Fig10}(b)]. The results therefore support the discussion of Sections   \ref{section3} and \ref{section4} on the role of unconventional magnetism for inducing a spin-singlet to spin-triplet conversion and also for changing the parity of the emergent superconducting correlations.

\bibliography{biblio}

\end{document}